%% file: main.tex
\documentclass[11pt]{article}

\usepackage{amsthm,amsmath,amsfonts,amssymb}
\usepackage[numbers,sort&compress]{natbib}
\usepackage[colorlinks=true,citecolor=blue,urlcolor=blue,linkcolor=blue]{hyperref}
\usepackage{graphicx}
\usepackage{xcolor}
\usepackage{underscore}
\usepackage{algorithm}
\usepackage{algpseudocode}
\usepackage{subfig}
\usepackage{comment}

\usepackage{float}

\theoremstyle{plain}

\theoremstyle{definition}

\title{Detecting Conflicts in Evidence Synthesis Models Using Score Discrepancies}

\author{
  Fuming Yang\thanks{MRC Biostatistics Unit, University of Cambridge, East Forvie Site, Robinson Way, Cambridge CB2 0SR, United Kingdom. Email: \texttt{fuming.yang@mrc-bsu.cam.ac.uk}} \and
  David J. Nott\thanks{Department of Statistics and Data Science, Institute of Operations Research and Analytics, National University of Singapore, 117546, Singapore. Email: \texttt{standj@nus.edu.sg}} \and
  Anne M. Presanis\thanks{MRC Biostatistics Unit, University of Cambridge, East Forvie Site, Robinson Way, Cambridge CB2 0SR, United Kingdom. Corresponding author. Email: \texttt{anne.presanis@mrc-bsu.cam.ac.uk}}
}

\date{}

\begin{document}
\maketitle

\begin{abstract}
Evidence synthesis models combine multiple data sources to estimate latent quantities of interest, enabling reliable inference on parameters that are difficult to measure directly. However, shared parameters across data sources can induce conflicts both among the data and with the assumed model structure. Detecting and quantifying such conflicts remains a challenge in model criticism. Here we propose a general framework for conflict detection in evidence synthesis models based on score discrepancies, extending prior–data conflict diagnostics to more general conflict checks in the latent space of hierarchical models. Simulation studies in an exchangeable model demonstrate that the proposed approach effectively detects between-data inconsistencies. Application to an influenza severity model illustrates its use, complementary to traditional deviance-based diagnostics, in complex real-world hierarchical settings. The proposed framework thus provides a flexible and broadly applicable tool for consistency assessment in Bayesian evidence synthesis.
\end{abstract}

\textbf{Keywords:} Bayesian, conflict, evidence synthesis, model criticism, score discrepancy


\input{Section1_Intro}
\input{Section2_Motivation}
\input{Section3_Method}

\input{Section4_Example}

\input{Section5_Flu_Example}

\input{Section6_Further}
\appendix
\input{Appendix}
\section*{Acknowledgments}
We thank Prof Daniela De Angelis and Dr Christopher Jackson at the MRC Biostatistics Unit, University of Cambridge, for helpful discussions.

\section*{Funding}
FY and AMP are funded by the UK Medical Research Council programme MC\_UU\_00040/04.  
David Nott's research was supported by the Ministry of Education, Singapore, under the Academic Research Fund Tier 2 (MOE-T2EP20123-0009).

\bibliographystyle{apalike}
\bibliography{ref}

\end{document}

%% file: Section1_Intro.tex
\section{Introduction}

In various fields, there is often interest in quantifying variables that are not directly observable, i.e. \emph{latent variables}. However, knowledge of such latent quantities, such as disease severity or effectiveness of treatments, is necessary for informed decision-making and they can often be indirectly estimated by combining information from multiple data sources. \emph{Evidence synthesis models}, often Bayesian hierarchical models, provide us with a formal framework for integrating diverse data sources in a probabilistic model, by mathematically expressing both deterministic and stochastic relationships between observed and latent variables, thereby performing inference using both direct and indirect data \citep{welton2012, DeAngelisPresanis2019}. Applications of such models are found in health technology assessments \citep{welton2012, wheaton2023}, comparative effectiveness research and clinical trials \citep{Ade2006, verde2016, wade2025}, epidemiological studies \citep{Presanis2021HIV, Melanie2022, anderegg2024}, and ecological modelling \citep{Blowes2024}.

However, conflicts may arise between observed data or between the data and model assumptions, including priors, the assumed likelihood, and other model assumptions, perhaps due to unaccounted biases in some data sources \citep{Ade2006, Presanis2008, Presanis2013}. Inconsistencies among these sources can lead to misleading or biased inference. After detecting and quantifying the conflicts, we then need to resolve them, before being able to draw reliable conclusions. Identifying and reporting these conflicts is a prerequisite for trustworthy model development, data fusion, and subsequent decision-making processes.

Although methods to detect and measure conflicting evidence exist \citep[e.g.][]{Yuan2012,gasemyr2009extensions,Presanis2013,Nott2021,Seth2019,Gasemyr2019}, they each have limitations that have so far prevented them from widespread use. To address these limitations, we propose extending Bayesian prior-predictive diagnostics into the latent space of evidence synthesis models. Our approach has three key components: first, we use score-based discrepancies \citep{Nott2021}, which use model expansions to target particular types of misspecification; second, we implement checks using these discrepancies through cross-validatory node-splitting \citep{Presanis2013}; and third, we draw on model criticism techniques for latent space from \citep{Yuan2012} and \citep{Seth2019} to establish appropriate reference distributions for calibrating our checks in hierarchical models.
This approach is well suited to models with many latent parameters, offering targeted detection of misspecification and a natural link to model expansion when existing parameters fail to capture key discrepancies \citep{Gasemyr2019}.

Section~\ref{sec:Existing Work and Motivations} gives basic background on existing approaches for conflict detection in the Bayesian framework, explains their limitations and how existing ideas can be combined, and motivates our proposed method. Section~\ref{sec:Methods} introduces our proposed methodology and its practical implementation. In Section~\ref{sec:Experiment for checking conflict between different data sources}, we demonstrate the method in a simulation example. Section~\ref{sec:Influenza Example} then presents a real-world case study on influenza severity estimation. Finally, Section~\ref{sec:Discussions and Further Research Directions} concludes with a brief discussion.

%% file: Section2_Motivation.tex
\section{Existing Work and Motivations for Extensions}
\label{sec:Existing Work and Motivations}

\subsection{Bayesian Hierarchical Models}
\label{sec:Bayesian Hierarchical Models}

We consider Bayesian inference using a model $H$ with parameters $\theta$ to describe data $\mathbf{y}$. It is assumed that $\mathbf{y}$ has a distribution with density $p(\mathbf{y} \mid \boldsymbol{\theta}, H)$, and may include independent observations from multiple sources or groups. We specify a prior $p(\boldsymbol{\theta} \mid H)$ that represents information about $\boldsymbol{\theta}$ before observing the data. Bayes' theorem gives the posterior $p(\boldsymbol{\theta} \mid \mathbf{y}, H) \propto p(\mathbf{y} \mid \boldsymbol{\theta}, H)p(\boldsymbol{\theta} \mid H)$, combining the prior with the likelihood. The parameters $\boldsymbol{\theta}$ can include hyperparameters (which control distributions of other parameters), global parameters, and unit-specific parameters.

Models with hierarchical layers are called \emph{multi-level models}, where parameters are themselves given probability distributions governed by higher-level parameters. For instance, individual observations indexed by $j$ (\(j = 1, \dots, n_i\)) within group $i$ (\(i = 1, \dots, I\)) may follow:
\begin{equation}
Y_{ij} \mid \lambda_i, \boldsymbol{\phi} \sim p(Y_{ij} \mid \lambda_i, \boldsymbol{\phi}) \quad \text{and} \quad  \lambda_i \mid \boldsymbol{\psi} \sim p(\lambda_i \mid \boldsymbol{\psi})
\label{eqn: general exchangeable class}
\end{equation}
where $\boldsymbol{\psi}$ and $\boldsymbol{\phi}$ are hyperparameters with prior distribution $p(\boldsymbol{\psi}, \boldsymbol{\phi} \mid H)$. The prior for $\lambda=\{\lambda_i, i=1,\dots, I\}\mid \boldsymbol{\psi}$ is hierarchical
because it is defined conditionally on $\psi$ which is given a prior itself.
Parameters such as $\lambda_i$ are often called \emph{latent variables}, as they are not observed directly but inferred from the observed data.

\begin{figure}[htbp]
\centering
\includegraphics[width=0.6\linewidth]{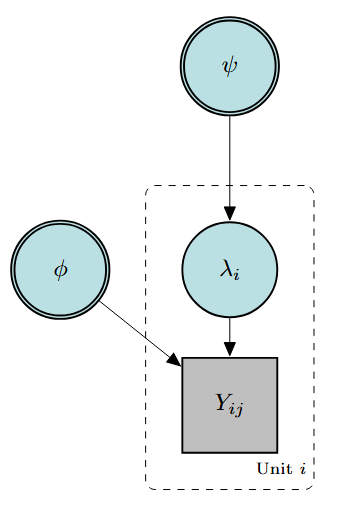}
\caption{An example of DAG showing the general exchangeable model class stated in Equation~\ref{eqn: general exchangeable class}. Double circles around $\boldsymbol{\psi}$ and $\boldsymbol{\phi}$ indicate founder nodes or hyperparameters, which are assigned prior distributions $p(\boldsymbol{\psi}, \boldsymbol{\phi} \mid H)$. Square nodes represent the observed data \(\mathbf{y}\), and solid arrows denote probabilistic (distributional) relationships, indicating that child nodes conditionally depend on parent nodes. Although not shown explicitly here, dashed arrows are typically used to represent deterministic or functional relationships between nodes. Repetition over units or within units (e.g.,~repeated measurements within groups) is typically indicated by dashed rectangles.}
\label{fig:general_DAG}
\end{figure}

Such models can usually be represented graphically by \emph{directed acyclic graphs} (DAGs) \citep{lauritzen1996}. Figure~\ref{fig:general_DAG} illustrates a general exchangeable model class as in Equation~\ref{eqn: general exchangeable class}, where observations within the same group are assumed to be \emph{exchangeable}, meaning their distribution does not depend on the ordering and dependence is captured through the shared group-level parameter $\lambda_i$. 

\subsection{Predictive Diagnostics}
\label{sec:Predictive Diagnostics}

To assess whether the observed data $\mathbf{y}$ are reasonably generated by the assumed model $H$, a natural idea is to compare an observed test statistic $T(\mathbf{y})$ to its reference distribution under the assumed model. Typical choices of $T$ are discrepancy statistics, functions of the data and parameters designed to detect specific components of misfit \citep{gelman2013bayesian}.  The comparison to a reference distribution allows us to evaluate whether the observed data appear extreme relative to the model. 


An initial implementation of this idea is given in \citep{Box1980} where the prior-predictive $p$-value for assessing model adequacy is introduced: 
$p=P(T(y^\text{rep}) \ge T(y^\text{obs}))$ and its two-sided version $
p^{(2)} = 2 \, \min \{ p, \, 1 - p \}
$
where $T$ is a discrepancy statistic and $T(y^\text{rep})$ is drawn from the prior-predictive distribution for $T$, having density $p(T(\mathbf{Y})) = \int p(T(\mathbf{Y}) \mid \boldsymbol{\theta}) \, p(\boldsymbol{\theta}) \, d \boldsymbol{\theta}$ for proper priors. The resulting $p$-value quantifies how \emph{surprising} the observed data are under the prior. 
{Evans and Moshonov} \citep{Evans2006} argued that prior-predictive diagnostics are appropriate for checking for conflict between the prior and likelihood. They further argued that $T$ should depend on the data only through a minimal sufficient statistic, since otherwise the prior-predictive check may depend on aspects of the data that are irrelevant to the likelihood, which would have nothing to do with prior-data conflict.  They also extended their approach to accommodate hierarchically specified priors. {Nott et al.} \citep{Nott2020} use the divergence between the prior and posterior as a discrepancy for a prior-data conflict check, employing Gaussian mixture variational approximations for tractable computation.  


To assess Bayesian \emph{model adequacy}, including both prior and likelihood components, posterior-predictive checks (PPCs) \citep{rubin1984bayesianly} remain the most widely used method. The approach compares the observed discrepancy measure \(  T(y^{\text{obs}}) \)  with the posterior predictive distribution of $T$ given the observed data, having density 
\(
p(T(\mathbf{Y}) \mid y^{\text{obs}}) = \int p(T(\mathbf{Y}) \mid \boldsymbol{\theta}) \, p(\boldsymbol{\theta} \mid y^{\text{obs}}) \, d \boldsymbol{\theta}.
\) Then we calculate a posterior-predictive $p$-value, defined analogously to the prior-predictive $p$-value. However, it is well known that the $p$-values produced by PPCs are not valid, in the sense that they are not uniformly distributed under correct model specification, even asymptotically \citep{gelman2013bayesian}. This leads to conservative checks, mainly because the data are used twice: once to generate posterior predictive distributions, and again to compute the observed discrepancy \citep{Bayarri2000, Bayarri2003, Bayarri2007}. 
To address the conservatism of posterior-predictive $p$-values, alternatives have been proposed that avoid the double use of data. One approach uses conditional and partial posterior-predictive $p$-values \citep{Bayarri2000, Robins2000, Bayarri2007}. These methods reduce the influence of the data on the discrepancy statistic when forming the predictive distribution. Another approach calibrates $p$-values to be uniformly distributed via a post-processing method involving double simulation \citep{hjort2006post, Gasemyr2019}.
However, these methods are computationally intensive.



The hierarchical nature of many models motivates the use of conflict detection within a \emph{mixed-predictive} framework \citep{Marshall2007}. This framework naturally extends standard predictive approaches, making them more suitable for handling \emph{latent-space} parameters and \emph{hierarchical structures}. The mixed predictive distribution of a discrepancy $T$ for a hierarchical model as in Figure~\ref{fig:general_DAG} has density
\(
p(T(\mathbf{Y}) \mid y^\text{obs}) \;=\;
\int p(T(\mathbf{Y}) \mid \boldsymbol{\lambda}, \boldsymbol{\phi})\,
     p(\boldsymbol{\lambda} \mid \boldsymbol{\psi})\,
     p(\boldsymbol{\phi}, \boldsymbol{\psi}\mid y^\text{obs})\,
     d\boldsymbol{\lambda}\, d\boldsymbol{\phi} \, d\boldsymbol{\psi}.
\) where $\boldsymbol{\lambda} = \{ \lambda_{i}, i = 1, \dots, I\}.$ This reference distribution is constructed by first drawing replicates of latent parameters $\boldsymbol{\lambda}$ from the marginal posterior of hyperparameters, and then integrating out these latent variables. The observed discrepancy statistic $T^\text{obs}$ is then compared to this reference distribution to test model/data compatibility. The resulting mixed-predictive $p$-value remains conservative, but typically less so than the posterior-predictive one. 

When multiple groups of observations are involved, both posterior- and mixed-predictive approaches are commonly embedded in a cross-validatory framework: the data are partitioned into a \emph{reference} set and a \emph{held-out (observed)} set. The reference set is used to update the model and generate the corresponding predictive distribution, while the held-out set provides the observed quantities against which this reference distribution is compared. By ensuring that the data used to build the predictive distribution are distinct from those used for evaluation, the framework avoids double use of the data and yields valid $p$-values under the null \citep{Marshall2007}, but this cross-validation can be computationally expensive \citep{gelman2013bayesian}.

\subsection{Conflict Detection in Evidence Synthesis Models}
\label{sec:Node-splitting Approach}



{Presanis et al.} \citep{Presanis2013}, in parallel with {Gåsemyr and Natvig} \citep{gasemyr2009extensions}, unifies and generalises the mixed-predictive framework \citep{Marshall2007} from the perspective of information contributions from different components, applying this node-splitting approach to evidence synthesis models. The key idea is to evaluate the {consistency of information contributing} to a specific node or edge in a Bayesian graphical model by isolating independent sources of evidence.

\begin{figure}[htbp]
\centering
\includegraphics[width=1\linewidth]{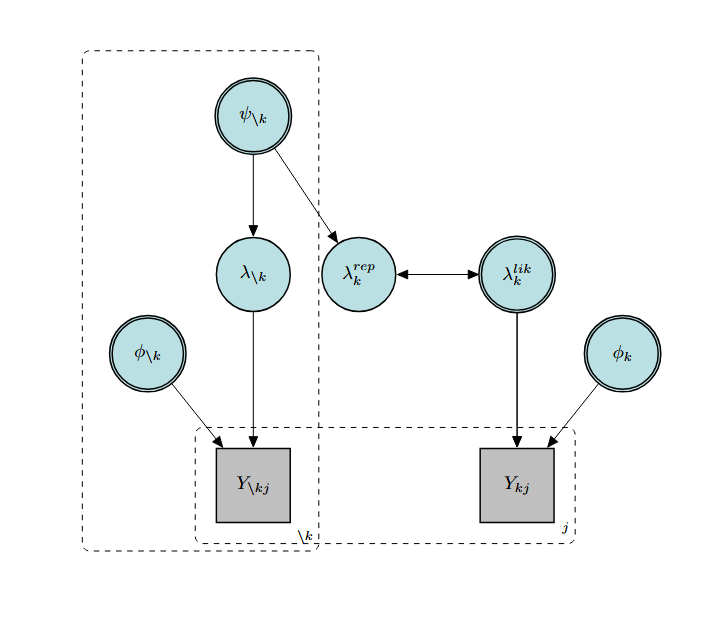}
\caption{A DAG where we split node $\lambda_k$ to assess conflict between information contributions. The long dashed box on the left provides indirect information about the parameter $\lambda_k$, containing data except $Y_k$ and corresponding parameters $\lambda_{\setminus  k}$, $\boldsymbol{\psi}_{\setminus  k}$, $\boldsymbol{\phi}_{\setminus  k}$ while data nodes $Y_{kj}$ provide direct information for $\lambda_k$. 
}
\label{fig:node-split_DAG}
\end{figure}

An illustrative example is shown in Figure~\ref{fig:node-split_DAG}, which depicts a general exchangeable model with the random effect node $\lambda_k$ selected for splitting. The contribution from $Y_{kj}$ informs the likelihood part of the split for $\lambda_k^{\mathrm{lik}} \sim p\big(\lambda_k \mid Y_k\big)$, while the predictive prior for $\lambda_k^{\mathrm{rep}} \sim p\big(\lambda_k \mid Y_{\setminus k}\big)$ is informed by the rest of the model excluding $Y_k$. The two resulting (independent) posterior distributions from the partitioned models are then compared to assess whether they lead to consistent inferences. 
The diagnostic is based on a \emph{conflict} $p$-value (similar to \citep{Marshall2007}): $
P^{(2)}_{k,\mathrm{con}}
= 2 \, \min \Big( P_{k,\mathrm{con}}, \; 1 - P_{k,\mathrm{con}} \Big),
$
where $ P_{k,\mathrm{con}} = \Pr\big(\lambda_k^{\mathrm{diff}} \le 0 \mid \mathbf{y}\big)$ and $
\lambda_k^{\mathrm{diff}} = \lambda_k^{\mathrm{rep}} - \lambda_k^{\mathrm{lik}},
$ which quantifies the extent to which the two sources of information disagree in their posterior inference.

{However, the interpretation of node-splitting results depends on the particular conflict measure used. For asymmetric distributions, one-sided versions of the conflict $p$-value may be appropriate, while for multimodal distributions a density-based $p$-value can be estimated empirically. However, such general measures of non-overlap may make the resulting notion of conflict less straightforward to interpret. More importantly for our purposes, although node-splitting can be designed to target particular locations in a model to detect conflict -- for example, to reflect differences between data and a model, between different data sources, or between a prior and a likelihood -- it does not exploit discrepancy statistics that could better characterize the nature of the conflict \citep{gelman2013bayesian}.}

The splitting process can typically be applied to any latent nodes in a DAG \citep{Presanis2013} and can be automated for all qualified splits in a network meta-analysis \citep{vanValkenhoef2016}. {Presanis et al.} \citep{presanis2017} explore this approach further with simultaneous hypothesis testing on groups of nodes and multiple partitions. These developments naturally suggest the possibility of detecting conflict between the local prior and the lifted likelihood at a given node, which in turn would allow prior–data conflict detection techniques to be applied using statistics defined only on latent parameters.


\subsection{Bayesian Model Criticism in Latent Space}
\label{sec:Model Criticism in Latent Space}

Another alternative for Bayesian model diagnostics in latent space arises from the fact that if parameters $\boldsymbol{\theta}$ are drawn from the prior, and data $\mathbf{Y}$ are generated from $p(\mathbf{Y} \mid \boldsymbol{\theta})$, then a \emph{single} posterior draw $\tilde{\boldsymbol{\theta}} \sim p(\boldsymbol{\theta} \mid \mathbf{Y})$ has a marginal distribution equal to the prior $p(\boldsymbol{\theta})$. 
This follows from writing $p(\boldsymbol{\theta},Y)=p(\boldsymbol{\theta})p(\mathbf{Y}\mid \boldsymbol{\theta})
=p(\mathbf{Y})p(\boldsymbol{\theta}\mid Y)$, so that drawing $\boldsymbol{\theta}$ then $Y$ sequentially from the joint Bayesian model is the same as drawing a single posterior sample for data drawn from the prior predictive.
Thus, if a posterior sample could not plausibly have been generated by the prior, then this indicates misspecification of some part of the model likelihood or prior.

We could directly compare posterior draws $\tilde{\boldsymbol{\theta}}$ with their marginal distribution when parameters are exchangeable within hierarchical structures, for instance using a goodness-of-fit test \citep{Seth2019}. A more general strategy is to consider \emph{pivotal quantities}, i.e. quantities with an invariant distribution under the true model \citep{degroot2011}, which provide a principled framework for assessing model adequacy by comparing observed quantities to reference distributions that are known under the assumed model. This approach avoids the difficulty that posterior samples derived from the same data are dependent, since pivotal quantities are parameter-free and follow a known distribution under the model. {Johnson} \citep{Johnson2007} introduces pivotal quantities based on both parameters and data as model diagnostics, while {Yuan and Johnson} \citep{Yuan2012} further propose that \emph{pivotal discrepancy measures} (PDM) can be pivotal quantities based solely on parameters, enabling diagnostics at latent levels. This flexibility is crucial for identifying structural conflicts beyond the data level. 

These methods utilise test quantities $T(\mathbf{Y},\boldsymbol{\theta})$ having a distribution that is invariant to the value of $\boldsymbol{\theta}$ under the assumed model, i.e. when $\mathbf{Y}$ is distributed according to $H$ with parameter $\boldsymbol{\theta}$. One then compares the observed value $T(y_\text{obs}, \tilde{\boldsymbol{\theta}})$, where $\tilde{\boldsymbol{\theta}}$ is a sample drawn from the posterior, to the known null distribution of $T$. Formal statements and theorems underpinning this approach are provided in {Appendix} \ref{Pivotal_theorems}. To illustrate, in the model of Figure~\ref{fig:general_DAG} with $\lambda_i \sim \mathcal{N}(\mu,\sigma^2)$, pivotal quantities can be written as $\tilde{z}_i=(\tilde{\lambda}_i-\tilde{\mu})/\tilde{\sigma}$, whose empirical distribution should match $\mathcal{N}(0,1)$ if the model is adequate.

For hierarchical models, the ``aggregated posterior checking'' framework proposed by {Seth et al.} \citep{Seth2019} emphasises the need to tailor reference distributions to the structure of the hierarchy. They propose that posterior samples of related latent variables that share the same prior distribution can be pooled and jointly compared against the corresponding conditional prior. For instance, to critique the prior \( p(\boldsymbol{\theta}_2 \mid \boldsymbol{\theta}_1) \), we wish to assess whether posterior samples $ \tilde{\boldsymbol{\theta}}_2 $ are in conflict with the conditional prior \( p(\boldsymbol{\theta}_2 \mid \tilde{\boldsymbol{\theta}}_1) \) conditioning on plausible posterior values of \( \boldsymbol{\theta}_1 \) as informed by the data.

So far, only a single posterior draw has been used to form an approximately pivotal quantity, but relying on a single draw rather than averaging over the full posterior introduces additional randomness into the model check.
A quantitative combination of dependent pivotal quantities can be obtained following \citep{Yuan2012}. Let realised pivotal discrepancies be
\(
T_g := T(y_\text{obs}, \tilde{\boldsymbol{\theta}}_g) \), \(g = 1, \dots, G,
\)
where $\tilde{\boldsymbol{\theta}}_g$ are posterior draws indexed by $g$. Marginally each $T_g$ has an invariant distribution $F$ if $y_\text{obs}$ is drawn from the prior predictive. Denoting $T_{(g)}$ as the $g$-th order statistic, an upper bound on the tail probability under dependence is
\[
P(T_{(g)} > t) \le \min \Big\{ 1, \frac{G(1 - F(t))}{G - g + 1} \Big\}.
\] 
Rather than fixing a particular order \(g\), {Yuan and Johnson} \citep{Yuan2012} proposed searching across all possible values of \(g\) to find the minimum \(p\)-value, \(p_{\min}\), while excluding extreme order statistics when the pivotal reference distribution is not exact. They interpreted \( p_{\min} < 0.25 \) as indicative of some evidence of conflict due to conservatism of the bound, and \( p_{\min} < 0.05 \) as strong evidence of conflict. 

{Liu et al. \cite{liu2025dependent} proposed an alternative approach to combining multiple dependent $p$-values in the context of integrating dependent studies.} They apply a heavy-tailed transformation to the individual $p$-values before combining them, and then compare the resulting statistic with its null distribution. They propose the Half-Cauchy Combination Test (HCCT) as a special case of approaches to reduce sensitivity to large p-values \citep{fang2023heavy,gui2025aggregating},  defining the combination statistic as:
\[
T_{\text{HCCT}} = \sum_{j=1}^{G} w_j F^{-1}_{\text{HC}}(1 - p_j) = \sum_{j=1}^{G} w_j \cot \Big( \frac{p_j \pi}{2} \Big),
\]
where \( \{p_1, \ldots, p_G\} \) are the \(p\)-values from dependent studies, and \( \{w_1, \ldots, w_G\} \), $\sum_{i = 1}^{G} w_i = 1$ are the corresponding weights to allow for unequal importance. Under mild dependence assumptions, they demonstrated that \( T_{\text{HCCT}} \) exhibits tail behaviour similar to that of a Half-Cauchy distribution. Although the exact density of $T_{\text{HCCT}}$ can be derived
under independence, {Liu et al.} \citep{liu2025dependent} showed that when the number of studies is large (e.g., \( G \geq 1000 \)), the distribution can be well approximated by the standard Landau distribution: 
\[
\text{Landau}\left( \frac{2}{\pi} \left( -\sum_{j=1}^{G} w_j \ln w_j + 1 - \gamma \right),\ 1 \right) 
\]
where \( \gamma \approx 0.5772 \) is the Euler–Mascheroni constant. 

{Covington and Miller} \citep{covington2025} employed a related approach to aggregate dependent $p$-values in Bayesian model checking based on the Cauchy combination test (CCT) \citep{liu2020cauchy}. Their strategy was to reparametrise the random elements of interest in the model into independent uniform random variables, perform tests targeting potential misspecification, and then combine the dependent test results.

A notable limitation of the latent space model criticism approach of \citep{Yuan2012} and \citep{Seth2019} is that it has not yet been extended to the context of multiple data sources.




\subsection{Score-Based Conflict Checks}
\label{sec:score-Based Conflict Checks}

Next we outline some score-based prior-data conflict checks considered in \citep{Nott2021}, and describe how to use this approach in conjunction with the methods of \citep{Johnson2007, Yuan2012, Seth2019} and/or \citep{liu2025dependent} for conflict detection at latent nodes in evidence synthesis models. A common way to assess models is through expansions that represent plausible departures from the assumed model, and by comparing the expanded model with the original one to judge whether the latter is adequate. In the Bayesian setting, however, an additional concern is prior–data conflict: {Nott et al.} \citep{Nott2021} therefore applied this parameter expansion notion to priors via score-type statistics, naturally extending classical score tests for checking likelihood. By introducing an auxiliary parameter $\alpha$ into the prior to generate a family of priors, one can design discrepancies that are sensitive to different forms of conflict, with possible choices discussed in Section~\ref{sec:choice of expansion}. The score statistic is defined as:
\begin{equation}
S_{\alpha}(\mathbf{y}) = \left. \frac{d}{d\alpha} \log p(\mathbf{y} \mid \alpha) \right|_{\alpha = \alpha_0} \label{eqn: score_discrepancy}
\end{equation}
where \( p(\mathbf{y} \mid \alpha) = \int p(\mathbf{y} \mid \boldsymbol{\theta}) p(\boldsymbol{\theta} \mid \alpha) \, d\boldsymbol{\theta} \) is the marginal likelihood. The original prior $p( \boldsymbol{\theta}) = p( \boldsymbol{\theta} \mid \alpha_0) $ is obtained when $\alpha = \alpha_0$. Here $\alpha$ is not treated as a hyperparameter, but rather chosen to detect specific aspects of conflict. 

As for prior-predictive diagnostics, the (one-sided) \emph{score-based $p$-value} is defined as $p_S = P(S_{\alpha}(y^\text{rep}) \geq S_{\alpha}(y^{\text{obs}}))$ where $y^\text{rep}$ follows the prior-predictive distribution $p(\mathbf{y} \mid \alpha)$ and $y^{\text{obs}}$ is the observation. Suppose $S_\alpha(\mathbf{y})$, $\mathbf{y} \sim p(\mathbf{y})$, is continuous, and $y^\text{obs} \sim p(\mathbf{y})$, then the tail probability is just one minus the transformation of $S_\alpha(y^\text{obs})$ by its distribution function, and hence is uniformly distributed on $[0, 1]$. So the $p$-value is a useful measure of surprise in the sense that we know what to expect from it if the data are generated under the prior predictive distribution. {For prior-data conflict diagnostics, this quantity may be evaluated directly. Our aim, however, is not merely to provide a Monte Carlo approximation to $p_S$, but to extend the score-based construction to latent components of hierarchical and evidence synthesis models, where the relevant uncertainty is propagated through posterior draws.}




It is useful to note that $S_{\alpha}(\mathbf{y})$ depends only on the data through the value of a minimal sufficient statistic, and it is invariant to the choice of that statistic, satisfying requirements for discrepancies set out by \citep{Evans2006}. 


\subsubsection*{Alternative Formulation}
\label{sec:Alternative Formulation}
An alternative version of the statistic, obtained via Fisher's identity under appropriate regularity conditions, can be written as
\begin{align}
S_{\alpha}(\mathbf{y}) &= \int \left. \frac{d}{d\alpha} \log p(\boldsymbol{\theta}\mid\alpha) \right|_{\alpha = \alpha_0} p(\boldsymbol{\theta}\mid \mathbf{y}) \, d \boldsymbol{\theta} \nonumber \\
&= \mathbb{E}_{\boldsymbol{\theta} \mid \mathbf{y}} \left( \left. \frac{d}{d\alpha} \log p(\boldsymbol{\theta}\mid\alpha) \right|_{\alpha = \alpha_0} \right)\label{eqn: alt_form}
\end{align} \citep{Nott2021}.
Here $S_{\alpha}(\mathbf{y})$ is the posterior expected ($\mathbb{E}_{\boldsymbol{\theta} \mid \mathbf{y}}$) rate of change of the log-prior with respect to the expansion parameter $\alpha$. This formula, together with its connection to prior-to-posterior divergence checks, gives an intuitive explanation that $S_{\alpha}(\mathbf{y})$ is large in magnitude if the posterior is concentrated out in the tails of the current prior and posterior samples can be made more plausible relative to the prior by changing $\alpha$, thus indicating a potential conflict. In its alternative form, the \emph{score discrepancy} can be considered as a function of posterior parameters $\boldsymbol{\theta} \mid \mathbf{y}$, depending only indirectly on data. 


This alternative form is therefore more convenient than the original for use in practice, since it can be easily approximated from a single posterior draw obtained by standard MCMC: given a draw \( \tilde{\boldsymbol{\theta}} \sim p( \boldsymbol{\theta} \mid y_{\text{obs}}) \), we obtain the \emph{randomised score discrepancy}:
\begin{equation}
S_{\alpha}(\tilde{\boldsymbol{\theta}}) = \left. \frac{d}{d\alpha} \log p(\tilde{\boldsymbol{\theta}} \mid \alpha) \right|_{\alpha = \alpha_0}
\label{eqn: randomised_score-discrepancy}
\end{equation}
where \( \alpha \) is an expansion parameter as in Equation~\ref{eqn: score_discrepancy}. Considering the framework in Section~\ref{sec:Model Criticism in Latent Space}, suitable reference distributions can be constructed also in a straightforward way: the randomised score discrepancy can be directly compared to the distribution of score values \( S_{\alpha}(\boldsymbol{\theta}) \), where \( \boldsymbol{\theta} \sim p(\boldsymbol{\theta}) \), to assess its extremity. The extremity can be calibrated by Monte Carlo simulation in practice, giving us a \emph{randomised score p-value}. 



With a single posterior sample, we implicitly assume the true $\boldsymbol{\theta}$ equals the sampled value. To relax this assumption and propagate uncertainty, we must consider multiple posterior samples and hence the joint distribution of score discrepancies. However, these randomised score discrepancies, being based on posterior draws, are neither independent nor pivotal. While it is often possible to construct asymptotically pivotal quantities from score statistics if the prior has suitable structure, which could be used within the framework of \citep{Yuan2012}, we instead adopt a different strategy: using the discrepancy in Equation~\ref{eqn: randomised_score-discrepancy} and generating reference distributions directly by simulating from the prior (or a conditional prior in hierarchical cases). This approach to a prior predictive check yields $p$-values with known marginal uniform distribution under the correct model, which can be regarded as pivotal discrepancies and therefore incorporated into \citep{Yuan2012}'s procedure for combining information across MCMC samples, or alternatively the \citep{liu2025dependent} approach to combining dependent studies (Section \ref{sec:Model Criticism in Latent Space}).

\subsubsection*{Hierarchical Extension}

{Nott et al. \citep{Nott2021} further extend the score-based prior--data conflict check to hierarchically specified priors. Consider a hierarchical model for $(\mathbf{y},\boldsymbol{\theta})$, with prior $p(\boldsymbol{\theta}) = p(\boldsymbol{\theta}_1) p(\boldsymbol{\theta}_2 \mid \boldsymbol{\theta}_1)$, where we want to check the consistency of $p(\boldsymbol{\theta}_2 \mid \boldsymbol{\theta}_1)$ with the observation for values of
$\boldsymbol{\theta}_1$ that are plausible under ${y}^{\mathrm{obs}}$. Again, ${y}^{\mathrm{obs}}$ denotes the observed data, whereas
${y}^{\mathrm{rep}}$ denotes a replicated dataset.} Following the mixed-predictive principles suggested by {Marshall and Spiegelhalter} \citep{Marshall2007}, {Nott et al.} \citep{Nott2021} suggest to generate predictive replicates and conduct the check as if $\boldsymbol{\theta}_1$ is fixed. Note that checking the conflict between $p(\boldsymbol{\theta}_1)$ and the data is the same as in the non-hierarchical case.

We consider an extension of the conditional prior: $p(\boldsymbol{\theta}_2 \mid \boldsymbol{\theta}_1, \alpha^{(1)})$ where the original prior $p(\boldsymbol{\theta}_2 \mid \boldsymbol{\theta}_1)$ corresponds to $\alpha^{(1)} = \alpha^{(1)}_0$. {We first consider an appropriate check of $p(\boldsymbol{\theta}_2\mid\boldsymbol{\theta}_1)$ as if $\boldsymbol{\theta}_1$ were known or fixed,}
\begin{align}
S_{\alpha}^{(1)}(\mathbf{y}, \boldsymbol{\theta}_1) 
&= \left. \frac{d}{d\alpha^{(1)}} \log p(\mathbf{y} \mid \boldsymbol{\theta}_1, \alpha^{(1)}) \right|_{\alpha^{(1)} = \alpha^{(1)}_0} \\
&= \mathbb{E}_{\boldsymbol{\theta}_2 \mid \mathbf{y}, \boldsymbol{\theta}_1} \left\{ \left. \frac{d}{d\alpha^{(1)}} \log p( \boldsymbol{\theta}_2 \mid \boldsymbol{\theta}_1, \alpha^{(1)}) \right|_{\alpha^{(1)} = \alpha^{(1)}_0} \right\}
\end{align}
where
$
p(\mathbf{y} \mid \boldsymbol{\theta}_1, \alpha^{(1)}) = \int p(\mathbf{y} \mid \boldsymbol{\theta}_1, \boldsymbol{\theta}_2) \, p(\boldsymbol{\theta}_2 \mid \boldsymbol{\theta}_1, \alpha^{(1)}) \, d\boldsymbol{\theta}_2.
$

{To account for the unknown $\boldsymbol{\theta}_1$, we then take its expectation with respect to the posterior distribution}
$p(\boldsymbol{\theta}_1\mid {y}^{\mathrm{obs}})$, defining
\begin{equation}
S_{\alpha}^{(1)}(\mathbf{y}) = \mathbb{E}_{\boldsymbol{\theta}_1 \mid y^{\text{obs}}} \left( S^{(1)}(\mathbf{y}, \boldsymbol{\theta}_1) \right).
\end{equation} We compare \( S_{\alpha}^{(1)}(y^{\text{obs}}) \) with \( S_{\alpha}^{(1)}(y^{\text{rep}}) \), where
$
y^{\text{rep}} \sim m(\mathbf{y}) = \int p(\boldsymbol{\theta}_2 \mid \boldsymbol{\theta}_1) \, p(\boldsymbol{\theta}_1 \mid y^{\text{obs}}) \, p(\mathbf{y} \mid \boldsymbol{\theta}) \, d\boldsymbol{\theta}$. {The reference distribution \( S_{\alpha}^{(1)}(y^{\text{rep}}) \) thus reflects knowledge of $\boldsymbol{\theta}_1$ under ${y}^{\mathrm{obs}}$, but uses the conditional prior of $\boldsymbol{\theta}_2$ given $\boldsymbol{\theta}_1$ in generating predictive replicates.}

In general, it is recommended that conditional priors are to be checked first, followed by the checks for hyper-priors \citep{Evans2006, Nott2021}. Overall model checks assess the fit of the entire model but ignore its hierarchical structure, while hierarchical checks specifically test for inconsistencies between different levels of the hierarchy. This allows hierarchical checks to detect conflicts in specific directions, such as discrepancies between group-level priors and individual-level data.

As before, given posterior samples, we could compare the score discrepancy at any posterior draw \( (\tilde{\boldsymbol{\theta}}_1, \tilde{\boldsymbol{\theta}}_2) \):
\begin{equation}
S_{\alpha}(\tilde{\boldsymbol{\theta}}) = \left. \frac{d}{d\alpha} \log p(\tilde{\boldsymbol{\theta}}_2 \mid \tilde{\boldsymbol{\theta}}_1, \alpha^{(1)}) \right|_{\alpha^{(1)} = \alpha^{(1)}_0},
\label{eqn: hierarchical-randomised}
\end{equation}
to its replicated version \( S_\alpha(\boldsymbol{\theta}_2, \tilde{\boldsymbol{\theta}}_1) \), where $\boldsymbol{\theta}_1$ is fixed at plausible values $\tilde{\boldsymbol{\theta}}_1$ and \( \boldsymbol{\theta}_2 \) is simulated from \(  p(\boldsymbol{\theta}_2 \mid \tilde{\boldsymbol{\theta}}_1) \).  This comparison checks whether the conditional prior $p(\boldsymbol{\theta}_2 \mid \boldsymbol{\theta}_1)$ is consistent with the data. In complex multi-level models, this allows us to construct a predictive reference distribution for any selected node, extending the prior-predictive diagnostics to move beyond data-level checks. This facilitates an extension from detecting (hierarchical) prior–data conflict to more general forms of model criticism in the latent space of DAGs or Bayesian hierarchical models.

As in the non-hierarchical case, marginally uniform $p$-values resulting from Equation~\ref{eqn: hierarchical-randomised} could serve as pivotal quantities. Although the proposed statistics represent useful measures of conflict or model fit, {Nott et al.} \citep{Nott2020} pointed out that in hierarchical settings valid $p$-values are harder to guarantee, and exact uniformity for finite samples rarely holds for checks based on conditional priors.


{
The construction above represents a general hierarchical mixed-predictive check and does not require partitioning the observations. In evidence synthesis models, however, we will consider in Section~\ref{sec:Sequential Updating} a cross-validatory setting in which the held-out child data are excluded when forming
$\boldsymbol{\theta}_1\mid y_{pa}^{\mathrm{obs}}$. The parent data still enter both the child-model inference and the construction of the reference distribution indirectly through this parent posterior. This construction reduces reuse of the held-out data, and the resulting $p$-values are generally expected to be better calibrated.
Nott et al. \citep{Nott2020} also suggested dividing the likelihood into components representing different data sources for conflict assessment.
}

More systematic rules for model partitioning and information flow restriction have also been developed within the Bayesian framework \citep{Presanis2013,presanis2017,LiuRob2025}. Excluding one of the data sources, we obtain posteriors of the link parameters between partitions, which can then be considered as a local ``prior'' to be updated with the left out data to recover the full posterior. Then the discrepancy, either on latent parameters or test quantities, that arises from this sequential updating can be evaluated to quantify the influence of the omitted data source on inference.
{
Unlike in a standard prior--data conflict check, this local ``prior'' is itself a posterior distribution and is generally not analytically available. Consequently, the standard score-based check cannot in general be applied directly.
} We formalise this sequential analysis procedure in Section~\ref{sec:Sequential Updating}. This approach is well-suited for evidence synthesis models, where multiple data sources are integrated into a single coherent statistical framework.


While obtaining a single $p$-value to evaluate model fit or detect conflict is desirable, sampling multiple values from the posterior of a link parameter to serve as local priors yields a distribution of $p$-values rather than a single summary. To address this, we employ the methods introduced in Section~\ref{sec:Model Criticism in Latent Space} (see also implementation details in Section~\ref{sec:Combining dependent $p$-values}) to aggregate these results into a {combined conflict probability}, providing an interpretable summary measure of overall adequacy.





%% file: Section3_Method.tex
\section{Methods}
\label{sec:Methods}

\subsection{Sequential Updating}
\label{sec:Sequential Updating}

Inspired by the ideas of node-splitting and cross-validation, we develop a sequential analysis framework to assess conflict between a specific component (or data source) and the remainder of the model. The procedure, summarised in Algorithm~\ref{alg:sequential}, proceeds as follows: (i) the data and model components are partitioned into a child sub-model on the data source of interest {$y_{ch}^\mathrm{obs}$} and its associated latent nodes (i.e. the likelihood part in \citep{Presanis2013, scheel2024}), and a parent part that summarises information from the remaining data {$y_{pa}^\mathrm{obs}$}, providing prior replicates \citep{Presanis2013} or local prior information \citep{scheel2024} for the target component; (ii) inference is performed sequentially: we first fit the parent model (stage $1$), excluding the target group, and then fit the child model (stage $2$) conditional on the posterior of the link parameters $\boldsymbol{\theta}_1$ obtained from the parent model. These link parameters, usually the parent nodes of the child model or the splitting nodes in \citep{Presanis2013}, {differ from typical node-splitting approaches in that} they are no longer assigned non-informative priors but are treated as fixed quantities set to the posterior draws from the stage $1$ model. 

For example, in Figure~\ref{fig:node-split_DAG}, contents in the long-dashed box that provides indirect information about $\lambda_k$ can be regarded as the parent model, while the child model corresponds to the likelihood component that supplies direct information through the observed data $Y_{kj}$. Instead of directly comparing $\lambda^{\text{lik}}_k$ and $\lambda^{\text{rep}}_k$, link parameters $\boldsymbol{\psi}$ and $\boldsymbol{\phi}$ are used to transfer information from the parent model to the child model. Graphical illustrations of this process are shown in Figure~\ref{fig:seq_simulation} for the simulation example in Section~\ref{sec:Experiment for checking conflict between different data sources}.

\begin{algorithm}[]
\footnotesize
\caption{Sequential Conflict Detection with Score Discrepancies}
\label{alg:sequential}
\begin{algorithmic}[1]

\State \textbf{Input:} \parbox[t]{0.85\linewidth}{%
  Full model \( p(\boldsymbol{\theta} \mid \mathbf{y}) \propto p(\mathbf{y} \mid \boldsymbol{\theta}) p(\boldsymbol{\theta}) \) with observations partitioned into parent data
\({y}_{{pa}}^{\mathrm{obs}}\) and held-out child data
\({y}_{{ch}}^{\mathrm{obs}}\), and parameters
\(\boldsymbol{\theta}
=
(\boldsymbol{\theta}_1,\boldsymbol{\theta}_2)\),
where \(\boldsymbol{\theta}_1\) are parent model parameters and
\(\boldsymbol{\theta}_2\) are child model parameters.%
}

\State \textbf{Goal:} Assess conflict between the child component and the rest of the model

\Statex \textit{/* Step 1: Fit the parent model excluding the child component */}
\State Fit the parent model via MCMC
\State Obtain posterior draws \( \tilde{\boldsymbol{\theta}}_1^{(1)}, \dots, \tilde{\boldsymbol{\theta}}_1^{(M)} \sim p( \boldsymbol{\theta}_1 \mid {y}_{{pa}}^{\mathrm{obs}}) \)

\Statex \textit{/* Step 2: For each draw from the parent, fit the child model */}
\For{each \( m = 1, \dots, M \)}
    \State Fix \( \boldsymbol{\theta}_1 = \tilde{\boldsymbol{\theta}}_1^{(m)} \) and fit the child model via MCMC
    \State Obtain the posterior draw \( \tilde{\boldsymbol{\theta}}_2^{(m)} \sim p( \boldsymbol{\theta}_2 \mid {y}_{{ch}}^{\mathrm{obs}},  \tilde{\boldsymbol{\theta}}_1^{(m)}) \)
\EndFor

\Statex \textit{/* Step 3: Define an expansion model with parameter \( \alpha \) */}
\State Define expanded model \( p(\boldsymbol{\theta}_2 \mid \boldsymbol{\theta}_1, \alpha) \)
\State Choose expansion direction \( \alpha \) targeting suspected conflict

\Statex \textit{/* Step 4: Generate prior-predictive replicates and compute score discrepancies for each posterior pair */}
\For{each \( m = 1, \dots, M \)}
    \State Compute score discrepancy:
    \[
        S_\alpha^{(m)} = \left. \frac{d}{d\alpha} \log p(\tilde{\boldsymbol{\theta}}_2^{(m)} \mid \tilde{\boldsymbol{\theta}}_1^{(m)}, \alpha) \right|_{\alpha = \alpha_0}
    \]
    \For{each \( g = 1, \dots, G \)}
        \State Sample \( \boldsymbol{\theta}_2^{(m,g)} \sim p(\boldsymbol{\theta}_2 \mid \tilde{\boldsymbol{\theta}}_1^{(m)}, \alpha_0) \)
        \State Compute:
        \[
            S_\alpha^{\text{prior}, (m,g)} = \left. \frac{d}{d\alpha} \log p(\boldsymbol{\theta}_2^{(m,g)} \mid \tilde{\boldsymbol{\theta}}_1^{(m)}, \alpha) \right|_{\alpha = \alpha_0}
        \]
    \EndFor
\EndFor

\Statex \textit{/* Step 5: Compute $p$-values by empirical comparison */}
\For{each \( m = 1, \dots, M \)}
    \State Compute:
    \[
        p^{(m)} = \frac{1}{G} \sum_{g=1}^G \mathbb{I}\left(S_\alpha^{\text{prior}, (m,g)} \geq S_\alpha^{(m)}\right)
    \]
\EndFor

\Statex \textit{/* Step 6: Combine individual $p$-values into a summary measure of conflict */}
\State Combine \( \{p^{(m)}\} \) via:
\State \quad (a) Upper bound-based minimum \(p\)-value procedure \citep{Yuan2012}, or
\State \quad (b) Half-Cauchy Combination Test \citep{liu2025dependent}

\State \Return Combined conflict probability, indicating whether there is a potential conflict

\end{algorithmic}
\end{algorithm}

This formulation allows us to assess how the information provided by the parent model deviates from the contribution of the target data source, detecting the conflict between the local prior and the lifted likelihood. The flexible partitioning of the model enables detection of multiple types of conflict---e.g., between data sources, between data and model, and between prior and likelihood---and can be applied to any latent node in a complex DAG. We refer to this process as \emph{sequential updating}. 


\subsection{Choice of Expansion Parameters \& Interpretation}
\label{sec:choice of expansion}

The score-type check based on parameter expansion can be viewed within the framework of sensitivity analysis \citep{Roos2015,Kallioinen2024}. The central idea is to investigate whether perturbations of certain parameters induce substantial changes in the posterior. From the perspective of local sensitivity, if the score discrepancy is large in magnitude, the marginal likelihood (and hence the Bayesian analysis) is highly sensitive to perturbations in the prior family, indicating potential conflicts. Conversely, if the discrepancy statistic is small, then the prior and data are consistent in the sense that small perturbations in the expansion parameter do not change the fit much.  

Typical directions for expansions include the following:  
\begin{enumerate} 
    \item \textbf{Normal priors:} \label{bullet:normal-prior}
    For a normal prior $\mathcal{N}(\mu_0, \sigma_0^2 / \lambda_0)$, natural expansion parameters include the mean $\mu$, the variance scale $\sigma$ (testing whether the posterior is sensitive to how diffuse the prior is), and the precision factor $\lambda$. This links to the power-scaling approach \citep{Kallioinen2024}, where scaling the prior by $\alpha>0$ rescales the variance as
    $
    \mathcal{N}(x \mid \mu, \sigma)^{\alpha} 
    \propto \mathcal{N}(x \mid \mu, \alpha^{-1/2}\sigma).
    $

    \item \textbf{Tail-heaviness, sparsity, and weighting:} Expansion parameters can be introduced through mixture weights that control either the tail–heaviness or sparsity of priors, or the relative contribution of specific model components. For example, one can use the mixture weight between an informative prior and an alternative diffuse prior as the expansion parameter to test whether conclusions are robust to tail–heaviness.



    \item \textbf{Double exponential families:} Exponential family distributions can be extended with an extra dispersion parameter, forming the double exponential family \citep{Efron1986double}. Expanding on this dispersion parameter $\tau$ (originally $\tau = 1$) provides a direct way to assess whether over- or under-dispersion relative to the original exponential family is justified.
    The resulting score discrepancy (obtained at $\tau = 1$) reduces to the deviance of the original exponential family plus a constant. This is demonstrated for the Normal, Poisson, and Binomial families in Appendix \ref{appendix: double}. As an example, in the normal case where $y \sim \mathcal{N}(\mu, \sigma^2/n)$, the corresponding double exponential family is 
    $
    y \sim \mathcal{N}(\mu, \sigma^2 / (n \tau)),
    $
    with $\tau$ controlling dispersion. This double exponential construction is also equivalent to the power-scaling approach \citep{Kallioinen2024}, and can be viewed as a generalisation of the Normal case in bullet point~\ref{bullet:normal-prior} to the exponential family.

\end{enumerate}

\subsection{Constructing Reference Distributions}

Following sequential updating, to assess potential conflict between different components of our Bayesian hierarchical model, we introduce a parametric expansion of the conditional prior distribution \( g(\boldsymbol{\theta}_2 \mid \boldsymbol{\theta}_1, \alpha) \), where \( \boldsymbol{\theta}_1 \) denotes vectors of parameters from the parent model, \( \boldsymbol{\theta}_2 \) are parameters of the child group conditional on \( \boldsymbol{\theta}_1 \), and \( \alpha \) is an expansion parameter with \( \alpha = \alpha_0 \) corresponding to the original prior. The direction of expansion is chosen to target components suspected of inducing or alleviating conflict, as discussed in Section~\ref{sec:choice of expansion}.

For each posterior draw \( (\tilde{\boldsymbol{\theta}}_1^{(m)}, \tilde{\boldsymbol{\theta}}_2^{(m)}) \) for \( m = 1, \dots, M \), we compute a score-type discrepancy \( S_\alpha^{(m)} \) that measures the sensitivity of the log-prior to pertubations in \( \alpha \). To construct reference distributions (Section~\ref{sec:score-Based Conflict Checks}), we generate predictive replicates \( \boldsymbol{\theta}_2^{(m,g)} \sim g(\boldsymbol{\theta}_2 \mid \tilde{\boldsymbol{\theta}}_1^{(m)}, \alpha_0)\), where \(g = 1, \dots, G \) indexes the predictive replicates, and we compute the corresponding discrepancies \( S^{(m,g)}_{\alpha, \text{prior}} \). These reference discrepancies are then compared against the observed \( S_\alpha^{(m)} \) to yield an empirical \( p \)-value for each posterior draw, measuring the level of conflict.


\subsection{Combining Dependent $p$-values}
\label{sec:Combining dependent $p$-values}

Given multiple $p$-values (each marginally uniform), we want to combine them to produce a conflict test statistic with easy interpretation. It is challenging to propose a rule of thumb for detecting conflict between model components based on multiple dependent randomised score \( p \)-values. This aggregation must account for dependence between the $p$-values and uncertainty of latent parameters propagated through random posterior sampling (e.g., via MCMC).

To address this, we adopt the two combination methods described at the end of Section \ref{sec:Model Criticism in Latent Space} \citep{Yuan2012, liu2020cauchy} to produce an indicative {combined conflict probability}, summarising and reflecting the overall level of conflict. For the approach of \citep{Yuan2012}, we search across all possible order statistics to obtain $p_{\min}$. For the HCCT method, we compute $T_{\text{HCCT}}$ assuming equal weights and calculate the resulting {combined conflict probability} based on the upper tail of the Landau approximation. Small values of $p_{\text{HCCT}}$ indicate stronger evidence of conflict within the model. 


%% file: Section4_Example.tex
\section{Simulation Example}
\label{sec:Experiment for checking conflict between different data sources}

Consider a simple hierarchical model of the following format:
\begin{align*}
y_{ij} &\sim \mathcal{N}( \theta_i, {\gamma}), \quad i = 1, \dots, 5, \quad j = 1, \dots, 10, \\
\theta_i &\sim \mathcal{N}( \beta, 5^2), \quad \gamma \sim \text{inv}\Gamma(2, 2), \quad \beta \sim \mathcal{N}(0,5^2).
\end{align*}

We consider a setup with five groups \(i = 1, \ldots, 5\) of data, each containing $10$ individuals. Let \( y_i = (y_{i1}, \dots, y_{i10})^\top \) denote the observations for group \( i \). We want to check if the \( i \)-th unit data, \( y_i \), is in conflict with the information provided by the remaining groups, \( y_{-i} \), and the prior. We simulate observations \( y_{ij} \) under this set up to obtain a baseline dataset without conflict. 

\begin{figure}[htbp]
    \centering
    \includegraphics[width=1\linewidth]{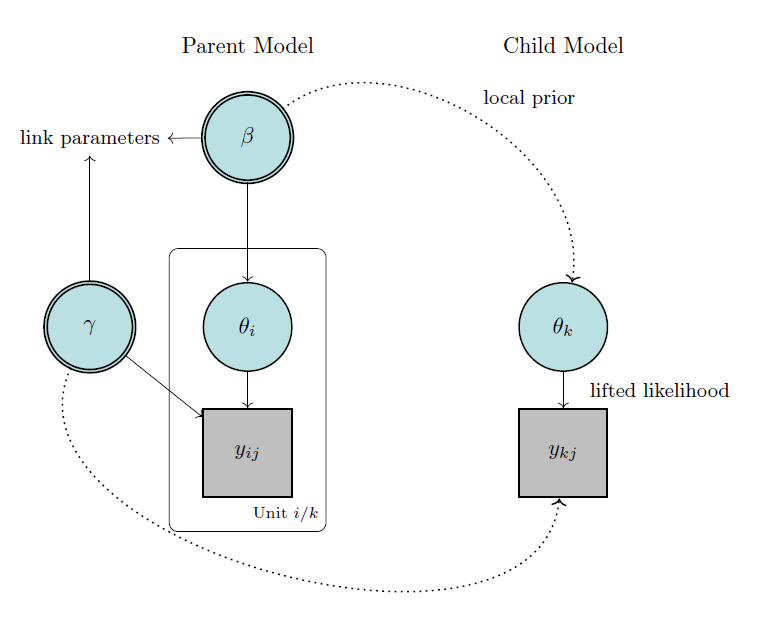}
    \caption{The sequential updating process of the simulation example. The separated group forms the child model, corresponding to the likelihood component, while the remainder constitutes the parent model. The link parameters $\beta$ and $\gamma$ are labeled, conveying local prior information to the child model.
}
    \label{fig:seq_simulation}
\end{figure}

To introduce conflict, we introduce a mean shift specifically to Group~$3$ in the simulation---setting \(\theta_3\) to a large value, e.g.,~\( \pm 20 \). We follow the procedure outlined in Algorithm~\ref{alg:sequential}, splitting out one group at a time. The link parameters are \( \beta \) and \( \gamma \), i.e.,~$\boldsymbol{\theta}_1 = (\beta, \gamma)$. We consider an expansion parameter $\alpha$ on $\theta_i \sim \mathcal{N}(\theta_i | \beta, 5\alpha)$. Figure~\ref{fig:seq_simulation} presents a graphical illustration.

\begin{figure*}[htbp]
\centering
\subfloat[No conflict between groups]{\label{fig:between_noconflict}
\includegraphics[width=0.8\linewidth]{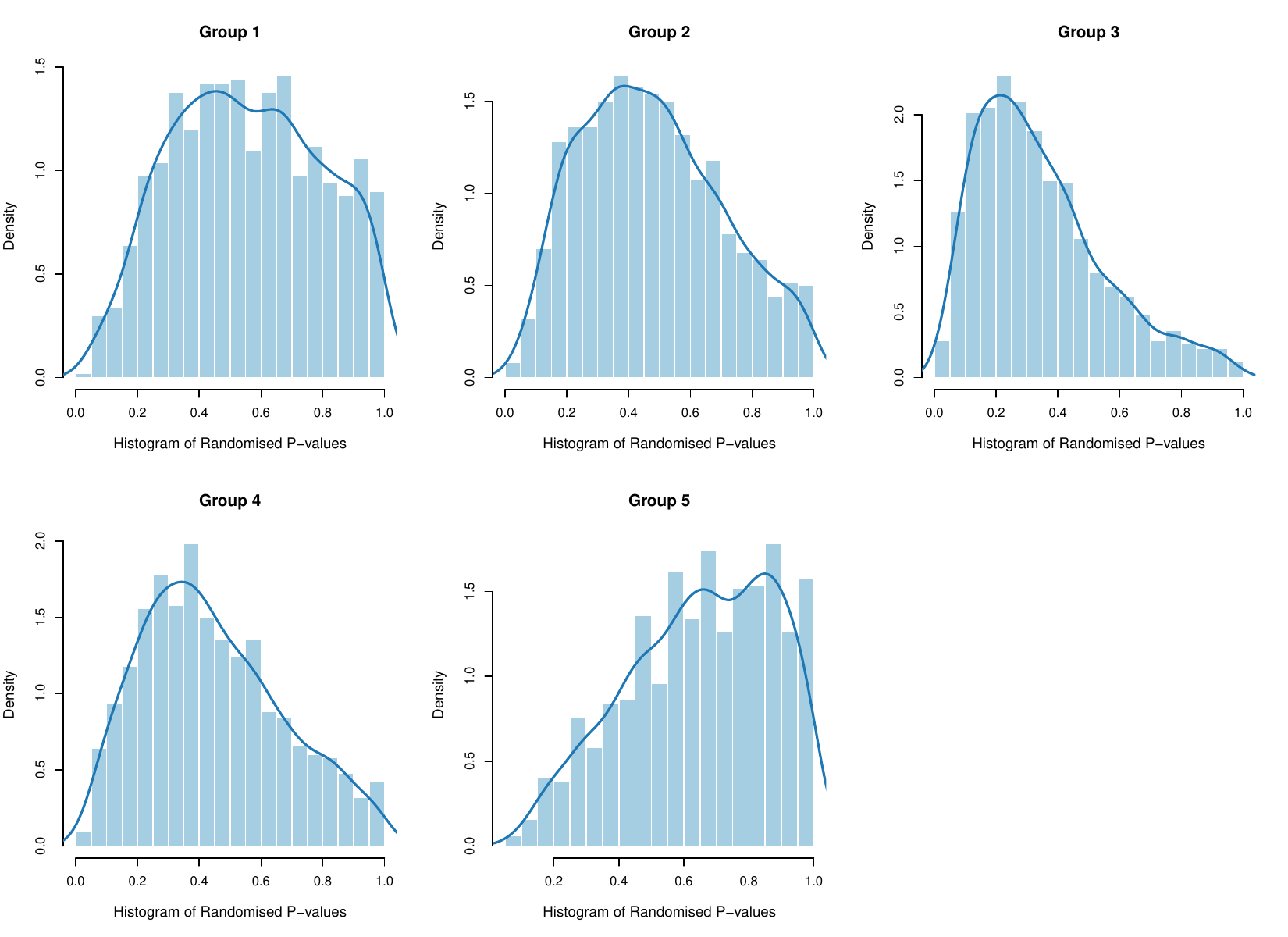}
}
\hfill
\subfloat[Conflict in Group $3$: $\theta_3 = 20$]{\label{fig:between_conflict}
\includegraphics[width=0.8\linewidth]{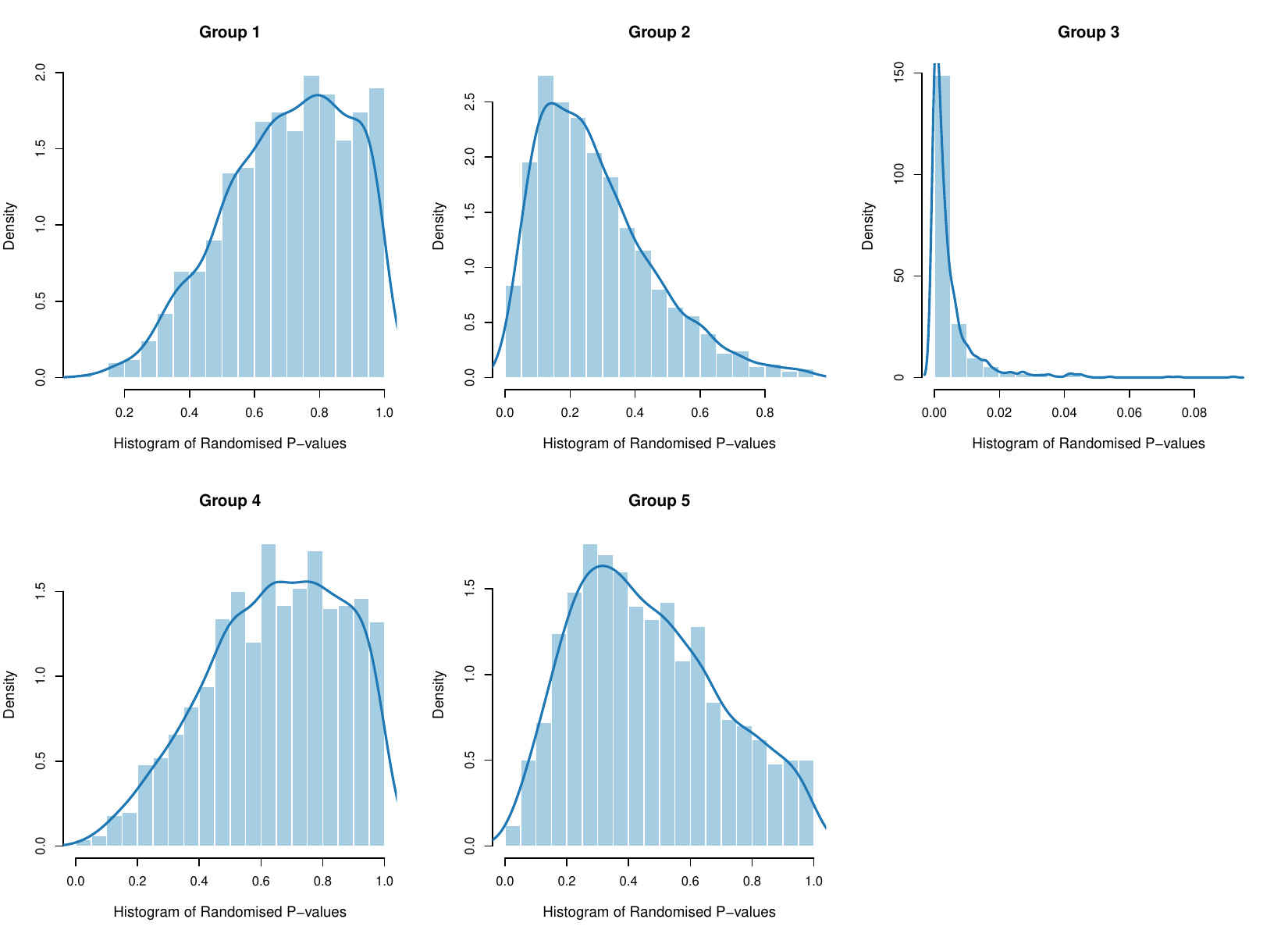}
}
\caption{The distribution of randomised $p$-values from the simulation example on detecting between-data conflict.}
\label{fig:between_group}
\end{figure*}

When a sufficiently large difference is introduced between Group~$3$ and the rest, the distribution of randomised \( p \)-values of each split reveals this inconsistency as shown in Figure~\ref{fig:between_conflict} (and Appendix \ref{appendix: splitting_simulated}). The randomised \( p \)-values for Group~$3$ shift toward $0$ as the mean difference increases, eventually becoming a sequence of zeros when the difference is large enough. Table~\ref{tab:between_group} shows that the combined {conflict probabilities} for the conflicting groups are small (although some do not fall below $0.05$, they are nevertheless close), while the values for other groups remain large. In addition, the Half-Cauchy combination tends to be more sensitive in detecting conflicts. For comparison, Figure~\ref{fig:between_noconflict} and Table~\ref{tab:between_group} illustrate the scenario without conflict, where all groups appear consistent.

For reference, we also include conflict $p$-values obtained from the node-splitting method. Since the two approaches quantify conflict in different ways (Appendix \ref{appendix: comparison}), their numerical results differ. Nevertheless, we observe that for the conflicting groups, the conflict 
$p$-values are close to $p_\text{min}$ and $p_\text{HCCT}$. The plots of $\theta_k^{\text{diff}}$ are provided in Appendix \ref{appendix: splitting_simulated}.

{This deliberately simple example is not intended to demonstrate uniform superiority over node-splitting, which performs well in this setting. Rather, it provides a transparent case in which the score-based diagnostics can be understood analytically and compared directly with an established conflict measure. The additional flexibility arises from the choice of expansion direction, which allows the diagnostic to target particular forms of model discrepancy.}

Building on this setup, we scale the simulation to $30$ groups, each with $50$ individuals. We inject conflict by setting \( \theta_3 = \theta_8 = \theta_{19} = 20 \). Results are summarised in Table~\ref{tab:between_large}, providing evidence of conflict in Groups~$3$, $8$, and $19$.


\begin{table}[htbp]
\centering
\begin{tabular}{lccccc}
\hline
\textbf{Group} & \textbf{1} & \textbf{2} & \textbf{3} & \textbf{4} & \textbf{5} \\
\hline
\multicolumn{6}{l}{\textbf{Conflict} $\theta_3 = 15$} \\
$p_\text{HCCT}$ & 1.000 & 1.000 & \textcolor{red}{0.020} & 1.000 & 1.000 \\
$p_{\min}$  & 0.995 & 0.648 & \textcolor{blue}{0.062} & 0.968 & 0.948 \\
Conflict $p$-value & 0.724 & 0.378 & \textcolor{blue}{0.057} & 0.604 & 0.574 \\
\hline
\multicolumn{6}{l}{\textbf{Conflict} $\theta_3 = 20$} \\
$p_\text{HCCT}$ & 1.000 & 0.857 & \textcolor{red}{0.003} & 1.000 & 1.000 \\
$p_{\min}$  & 0.997 & 0.474 & \textcolor{red}{0.004} & 0.994 & 0.822 \\
Conflict $p$-value & 0.855 & 0.28 & \textcolor{red}{0.004} & 0.718 & 0.459 \\
\hline
\multicolumn{6}{l}{\textbf{No Conflict}} \\
$p_\text{HCCT}$ & 1.000 & 1.000 & 0.998 & 1.000 & 1.000 \\
$p_{\min}$  & 0.978 & 0.836 & 0.586 & 0.779 & 0.997 \\
Conflict $p$-value & 0.579 & 0.486 & 0.351 & 0.456 & 0.712 \\
\hline
\end{tabular}
\caption{Summary of {combined conflict probabilities} from the simulation example on detecting between-data conflict. For comparison, conflict $p$-values obtained via the node-splitting method (Section~\ref{sec:Node-splitting Approach}) are also reported. Small values, indicating  evidence of conflict, are marked in red $(<0.05)$ and blue $(>0.05)$.}

\label{tab:between_group}
\end{table}

\begin{table*}[htbp]
\centering

\begin{tabular}{cccc|cccc}
\hline
\textbf{Group} & \textbf{\(p_{\min}\)} & \textbf{\(p_{\mathrm{HCCT}}\)} & Conflict $p$ &
& \textbf{\(p_{\min}\)} & \textbf{\(p_{\mathrm{HCCT}}\)} & Conflict $p$ \\
\hline
1 & 0.829 & 1.000 & 0.587 & 16 & 0.896 & 1.000 & 0.646 \\
2 & 0.141 * & 0.177 * & 0.093 & 17 & 0.213 & 0.469 & 0.134 \\
3 & 0.063 * & 0.042 ** & 0.038 ** & 18 & 0.997 & 1.000 & 0.834 \\
4 & 0.999 & 1.000 & 0.979 & 19 & 0.056 * & 0.036 ** & 0.035 ** \\
5 & 0.300 & 0.913 & 0.192 & 20 & 0.250 & 0.711 & 0.168 \\
6 & 0.447 & 1.000 & 0.292& 21 & 0.314 & 0.957 & 0.200 \\
7 & 0.141 * & 0.169 * & 0.087 & 22 & 0.998 & 1.000 & 0.804 \\
8 & 0.072 * & 0.052 ** & 0.043 ** & 23 & 0.999 & 1.000 & 0.876 \\
9 & 0.167 & 0.259 & 0.106 & 24 & 0.532 & 1.000 & 0.356 \\
10 & 0.528 & 1.000 & 0.344 & 25 & 0.998 & 1.000 & 0.800 \\
11 & 0.814 & 1.000 & 0.563 & 26 & 0.998 & 1.000 & 0.882 \\
12 & 0.619 & 1.000 & 0.428 & 27 & 0.207 & 0.455 & 0.134\\
13 & 0.905 & 1.000 & 0.645 & 28 & 0.869 & 1.000 & 0.622\\
14 & 0.991 & 1.000 & 0.744 & 29 & 0.196 * & 0.394 & 0.123\\
15 & 0.999 & 1.000 & 0.989 & 30 & 0.963 & 1.000 & 0.696 \\
\hline
\end{tabular}
\caption{Summary of {combined conflict probabilities} from the simulation example on detecting between-data conflict with $30$ groups and $50$ individuals each. {Smaller combined conflict probabilities indicate stronger evidence of conflict; values below $0.05$ are marked **, and those below $0.25$ are marked *.}}
\label{tab:between_large}
\end{table*}

%% file: Section5_Flu_Example.tex
\section{Influenza Example}
\label{sec:Influenza Example}

Application to an influenza severity model illustrates the use of our approach, complementary to traditional deviance-based diagnostics, in complex real-world hierarchical settings.

\subsection{Model and its Alternatives}

\begin{figure*}[htbp]
    \centering
    \includegraphics[width=\linewidth]{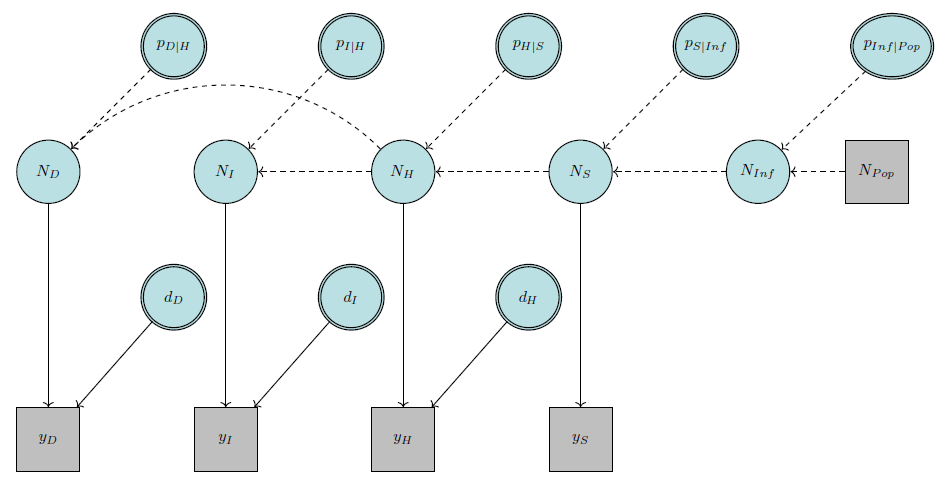}
    \caption{DAG of the flu model. In the DAG, Pop denotes all population, Inf denotes all infections, S denotes symptomatic infections, H denotes hospitalisations, and D denotes deaths. For brevity, we do not explicitly label the indices for different age groups in the DAG. $p_{\ell \mid l}$ represents the conditional probability in Equation~\ref{eqn: deterministic}. $N_{\ell}$ denotes the population size at severity level $\ell$, $y_{\ell}$ are the observed counts or estimates at level $\ell$, and $d_{\ell}$ is the detection probability.
}
    \label{fig:flu_DAG}
\end{figure*}

We consider a stochastic extension of the Bayesian model introduced in \citep{anne2014} (the corresponding DAG in Figure~\ref{fig:flu_DAG}) to assess the severity of the third wave of pandemic A/H1N1pdm influenza infections in the UK in winter 2009/2010. Monitoring and tracking the severity of an influenza outbreak is critical for guiding timely and proportionate public health responses, including determining the necessity of intervention measures. Severity of a disease outbreak is typically measured in terms of infection and case-severity risks, i.e. the probabilities that an infection or diagnosed infection (case) lead to a severe event such as consulting a GP, hospital or ICU admission or death. However, since infections---particularly mild or asymptomatic ones---are rarely directly observed, it is challenging to estimate infection-severity risks from a single data source. We typically only observe infections at the severe end of the ``severity pyramid'', such as ICU admissions or deaths.

Therefore, the most feasible approach involves integrating multiple data sources and accounting for their observational biases. Our analysis combines information on high-severity outcomes from several surveillance systems, thus a model is needed relating the severity process to these observations.


We model severity levels from the ``severity pyramid'' using a chain of conditional probabilities. The known severity levels, ordered from least to most severe, include: all infections (INF), symptomatic infections (SYM), GP consultations (GP), hospitalisations (HOS), ICU admissions (ICU), and deaths (DEA). We additionally use POP to denote the total population, with the population size $N_{a,\mathrm{POP}}$ for each age group $a$ assumed to be known. The severity model in \citep{anne2014} represents the number of influenza cases at a given severity level $\ell$ as a proportion of cases at the adjacent, less severe level $l$. This creates a ``nested'' structure across severity levels, where the proportions, or rather conditional probabilities $p_{a,\ell \mid l}$ for each age group $a$, are unknown parameters to be estimated, with prior Beta distributions. The population size $N_{a,\ell}$ for different age groups $a$ at each severity level $\ell$ is then modeled as a deterministic function of the conditional probabilities and the size at the preceding level, therefore also unknown and to be estimated, but without explicit prior distributions:
\begin{align}
p_{a, \ell \mid l} & \sim \text{Beta}(\alpha_{a,\ell},\beta_{a,\ell}) \nonumber \\
N_{a, \ell} & = p_{a, \ell \mid l} \cdot N_{a, l}
\label{eqn: deterministic}
\end{align}
where $\ell \in \{\text{INF}, \text{SYM}, \text{GP}, \text{ICU}, \text{DEA}\}$ and $l \in \{\text{POP},$ $ \text{INF}, \text{SYM}, \text{HOS}, \text{HOS}\}$. Here, $a$ indexes the five age groups into which the population is stratified: $0$–$4$, $5$–$14$, $15$–$44$, $45$–$64$, and $65+$ years. This structure, and the modeling of severity levels, is represented in the upper part of the DAG (Figure~\ref{fig:flu_DAG}). 


Informative Beta priors with known parameters $\alpha_{a,\ell},\beta_{a,\ell}$ are used in Equation~(\ref{eqn: deterministic}) to incorporate prior knowledge about the conditional probabilities at each severity level. These priors are either derived from posterior estimates based on earlier pandemic waves in 2009, or set to be weakly informative or flat in the absence of prior information. For further details, we refer the reader to the two-stage modeling framework described in \citep{anne2014}.


Focusing on the third wave, we model the number of infections at the lowest severity level with a uniform prior on the infection attack rate $p_{a,\text{INF} \mid \text{POP}}$ for each age group $a$:
\begin{align}
&p_{a,\text{INF} \mid \text{POP}}  \sim \text{Beta}(1, 1) \nonumber \\
&N_{a, \text{INF}}  = p_{a,\text{INF} \mid \text{POP}} \cdot N_{a, \text{POP}}
\end{align}

Our data to integrate are: GP consultation estimates derived from a previous model based on GP testing data; hospital admissions reported by a hospital sentinel system piloted in 2010/11; mortality data; and cumulative ICU admissions informed by an immigration-death submodel based on data on prevalent numbers of individuals in ICU with influenza. Full details of these data sources and submodels are available in \citep{anne2014, anne2009}. The posterior estimates from the submodels or observations typically represent a lower bound on the true number of cases at the corresponding severity level. These counts ($y$s in the DAG) contribute to the overall likelihood through binomial observation models, with probability parameter $d_\ell$ (assumed non age-specific) governing the probability of observing a case at level $\ell$ (a ``detection probability''). These observations or estimates, along with the detection probabilities, are represented in the lower part of the DAG (Figure~\ref{fig:flu_DAG}).

\subsubsection{Beta-Binomial modelling \label{sec: sec_bb}}

\begin{figure*}[htbp]
    \centering
    \includegraphics[width=\linewidth]{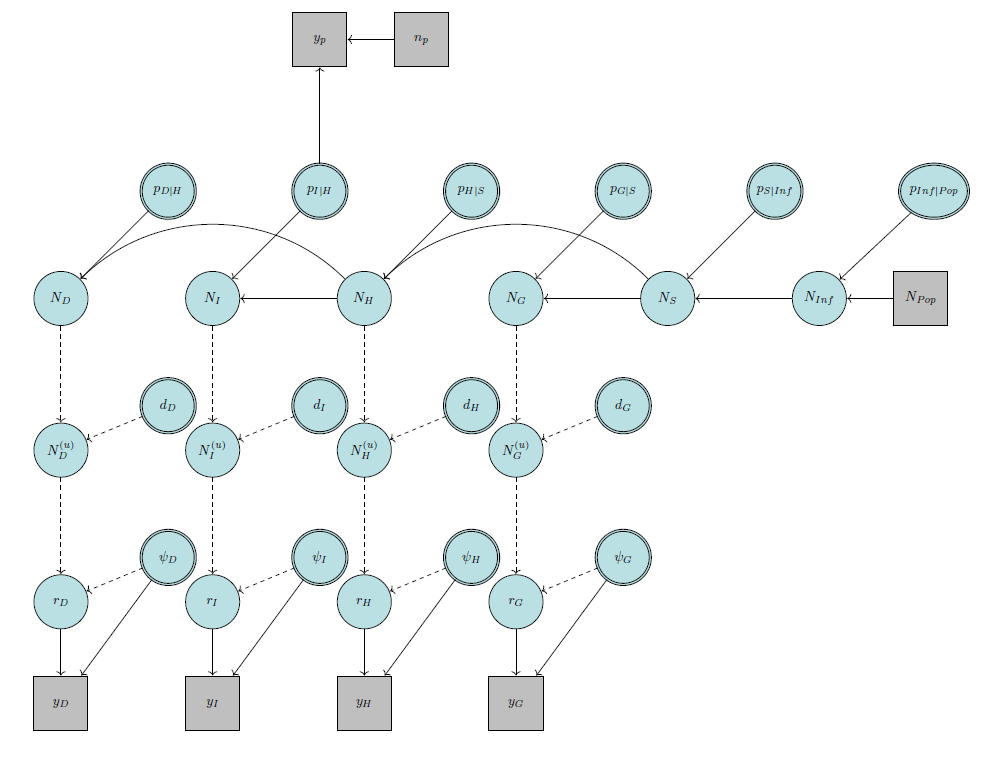}
    \caption{Stochastic extension of the flu model shown in Figure~\ref{fig:flu_DAG}. In addition to the notations used previously, G denotes GP concultations, $N^{(u)}_{\ell}$ represents the (potentially) under-ascertained number of infections at level $\ell$ in Equation~\ref{eqn:under-ascertainment}, $r_{\ell}$ is the size parameter in Equation~\ref{eqn:over-dispersion} and $\psi_{\ell}$ is the corresponding dispersion parameter.}
    \label{fig:flu_DAG_extension}
\end{figure*}

In the deterministic parameterisation of the model described above, the functional relationship between the numbers and the probabilities expresses an \emph{expected} number of infections at each severity level. However, in analogy to the literature on chain-binomial models \citep{Allen2008}, the relationship between the numbers $N_{a, \ell}$ and the conditional probabilities $p_{a, \ell \mid l}$ could instead be expressed stochastically, as a series of nested binomial random variables, i.e.,~a hierarchical binomial model of the form:
\begin{align}
p_{a, \ell \mid l} & \sim \text{Beta}(\alpha_{a,\ell},\beta_{a,\ell}) \nonumber \\
N_{a, \ell} & \sim \text{Binomial}(N_{a, l}, p_{a, \ell \mid l})
\label{eqn: Beta+Binomial}
\end{align} The relationship between severity levels is represented in the DAG shown in Figure~\ref{fig:flu_DAG_extension}.

The deterministic parameterisation can be viewed as a special case of this formulation, where $\mathbb{E}(N_{a,\ell}) = p_{a, \ell \mid l} N_{a,l}$. In contrast, the probabilistic (stochastic) transitions explicitly incorporate uncertainty between severity levels by accounting for chance events, which is particularly important when the number of infections is small—such as during the early stages of an epidemic or at high severity levels. Rather than treating the population at each severity level deterministically, this approach models individuals at a less severe level as having a probabilistic risk of progressing to more severe outcomes, offering greater flexibility and better aligning with our definition of case/infection-severity risk.

Moreover, the binomial relationship described above is itself a special case of the double binomial distribution introduced by {Efron} \citep{Efron1986double}, with the dispersion parameter $\tau = 1$. When additional flexibility is required -- specifically, a second parameter that allows the variance to be controlled independently of the mean, thereby accommodating over- or under-dispersion -- we can consider distributions from the double exponential family. Any conflict detected under the deterministic parameterisation may simply reflect a failure to capture uncertainty, rather than evidence of actual conflict. We therefore use the stochastic formulation to assess potential conflicts.


Note that many standard MCMC algorithms often struggle to efficiently sample from binomial distributions with large, unknown denominators, and this inefficiency can lead to poor chain mixing \citep{Zanella2020}. To avoid this issue, we can directly sample from the beta-binomial distribution:
\begin{align}
N_{a, \ell} & \sim \text{Beta-Binomial}(N_{a, l}, \alpha_{a,\ell},\beta_{a,\ell}) \nonumber\\
p_{a, \ell \mid l} & = N_{a, \ell} / N_{a, l}
\end{align}
This parameterisation is exactly equivalent to Equation (\ref{eqn: Beta+Binomial}) mathematically.

Apart from accounting for the relationships between severity levels, the model is amended in two additional aspects. First, a GP consultation level is introduced between the symptomatic infection and hospitalisation levels. This allows direct incorporation of GP consultation estimates and their uncertainty from the previous model, and also enables feedback from the severity model to GP-related parameters. Second, over-dispersion in the count data $y_{a,\ell}$ is accommodated via a negative binomial formulation $\text{Negative-Binomial}(\psi_{a,\ell}, r_{a,\ell})$. This improves model fit and mixing while addressing under-ascertainment and uncertainty from source models. Details of these extensions are given in Appendix \ref{Flu_extension}.

\subsection{Model Results and Model Criticism}
\label{criticism}

For inference, we use standard MCMC such as the algorithms implemented in JAGS, as Hamiltonian Monte Carlo (e.g.,~as implemented in Stan) does not support non-differentiable discrete parameters unless they are integrated out. 
Fitting the hierarchical binomial model and the beta-binomial model directly causes poor mixing due to inefficient sampling of binomial counts or MCMC chains jumping into areas of infinite log density. 
So we fit a hierarchical over-dispersion model in JAGS with deterministic relationship between severity levels, while introducing the stochastic relationship to check for conflict between data sources.

\begin{figure}[htbp]
\centering
\subfloat[Posterior number of infections by severity level and age]{\label{fig:rjags_output1}
\centering
\includegraphics[width=0.8\linewidth]{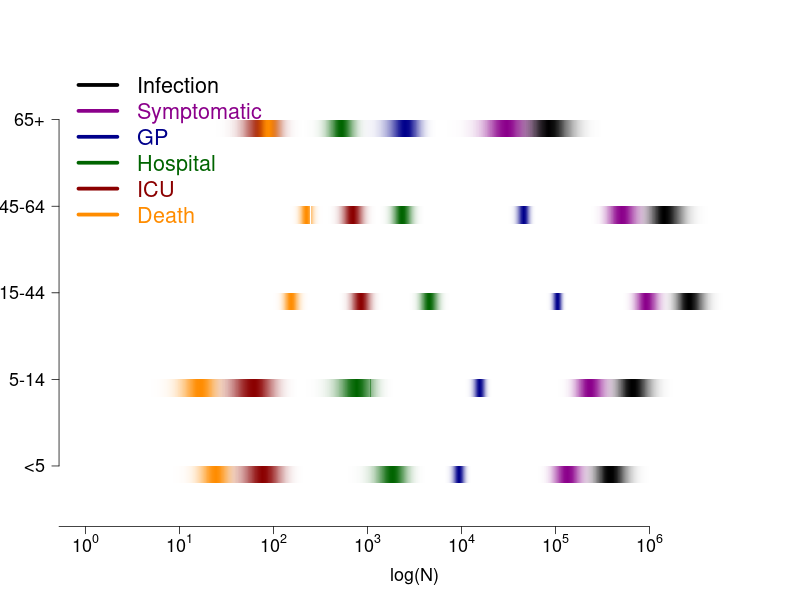}
}
\hfill
\subfloat[Posterior case-severity risks]{\label{fig:rjags_output2}
\centering
\includegraphics[width=0.8\linewidth]{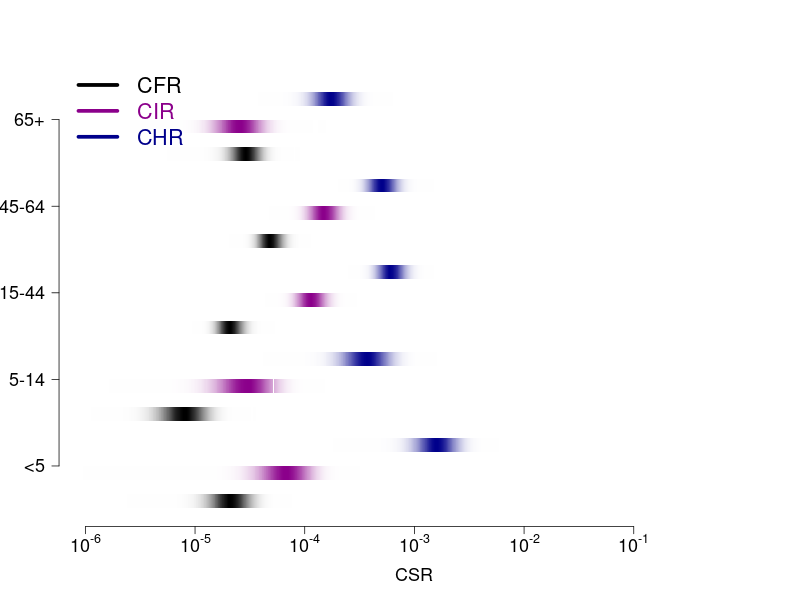}
}

\caption{Summary of posterior distributions of numbers of infections (top) and case-severity risks (bottom). The case-severity risks, defined as conditional probabilities, include death given infection (case-fatality risk, CFR), ICU admission given infection (case-ICU admission risk, CIR), and hospitalisation given infection (case-hospitalisation risk, CHR).
}
\label{fig:posterior}
\end{figure}

From our deterministic over-dispersion model, we confirm that the MCMC has converged by examining trace plots of the chains. Figure~\ref{fig:posterior} is the plot of posterior numbers of infections by severity level and age. We also compute the case-severity risks: the probabilities of severe events given either symptomatic or all infection, which are defined as functions of the conditional probabilities.  

As our initial stage of model criticism, we assess how well the model fits the data by looking at deviance summaries: there is some indication of slight lack of fit to the sentinel hospital data (in age groups $45-64$ and $65+$) on the proportion of hospitalisations leading to ICU admission ($p_{I \mid H}$), with Deviance Information Criterion (DIC) contributions greater than $1$: $3.82$ and $3.53$, respectively. In addition, we note that the node-splitting method mentioned in Section~\ref{sec:Node-splitting Approach} is challenging to apply here, as poor identifiability of the sub-models leads to multimodal posteriors.



\subsection{Results from Randomised Score-based Checks}

We follow the procedure outlined in Algorithm~\ref{alg:sequential}, isolating the data source associated with severity level $\ell$ to assess whether the data at this level conflict with data from other sources. Posterior samples for the preceding severity levels (i.e., $N'_l = N'_{\ell - 1}$ or $N'_l = N'_{\ell - 2}$) are obtained from the MCMC chains and used as linkage parameters $\boldsymbol{\theta}_1 = N_l$. To introduce the expansion parameter $\tau$, we consider extending the binomial relationship between severity levels to double-binomial \citep{Efron1986double}:
$
N_\ell \sim \text{Double-Binomial}(N_l, p_{\ell \mid l}, \tau),
$
where $\tau$ is a dispersion parameter.

\begin{figure*}[htbp]
\centering

\subfloat[]{\label{fig:obs_ICU}
\includegraphics[width=0.47\linewidth]{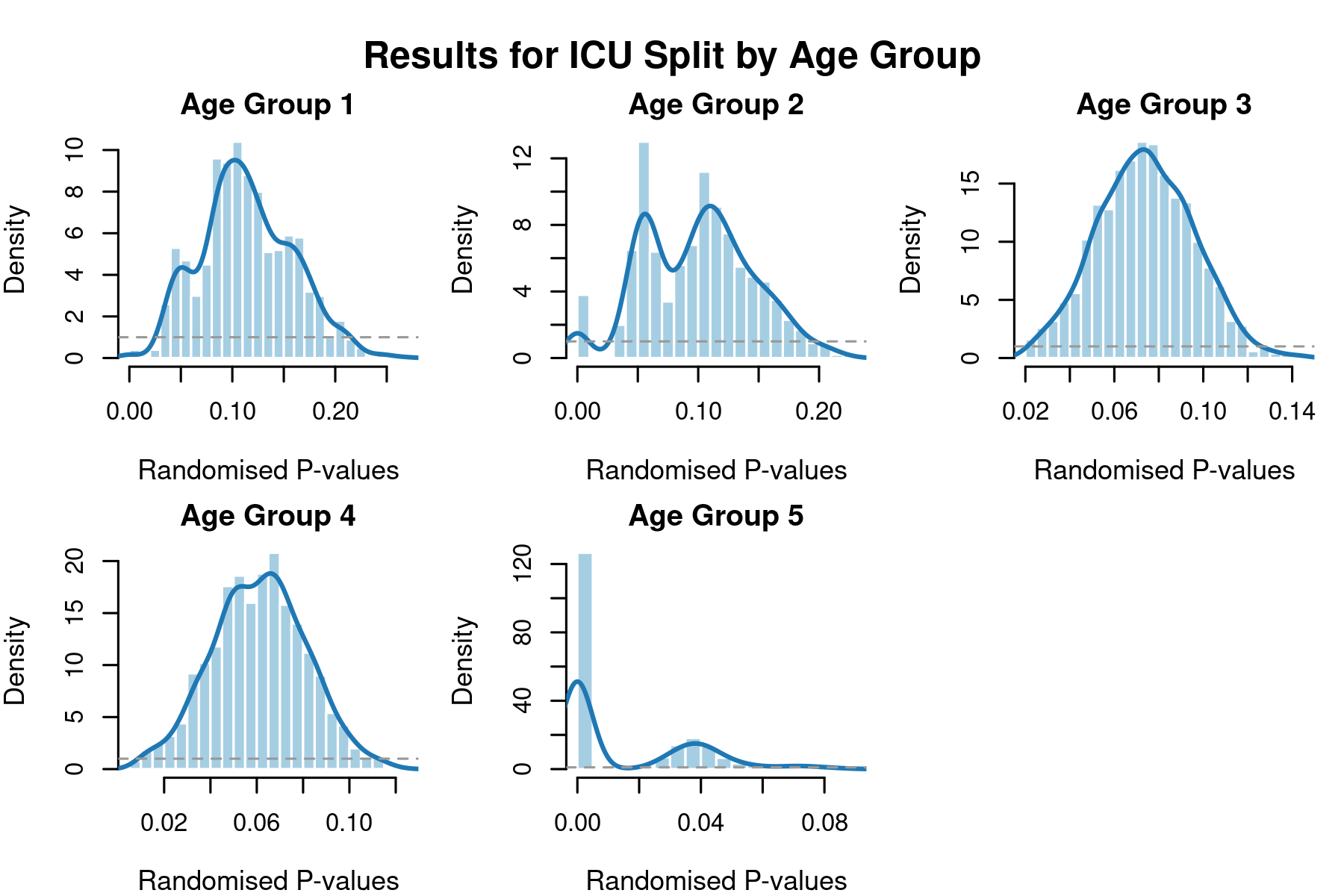}
}
\hfill
\subfloat[]{\label{fig:obs_DEA}
\includegraphics[width=0.47\linewidth]{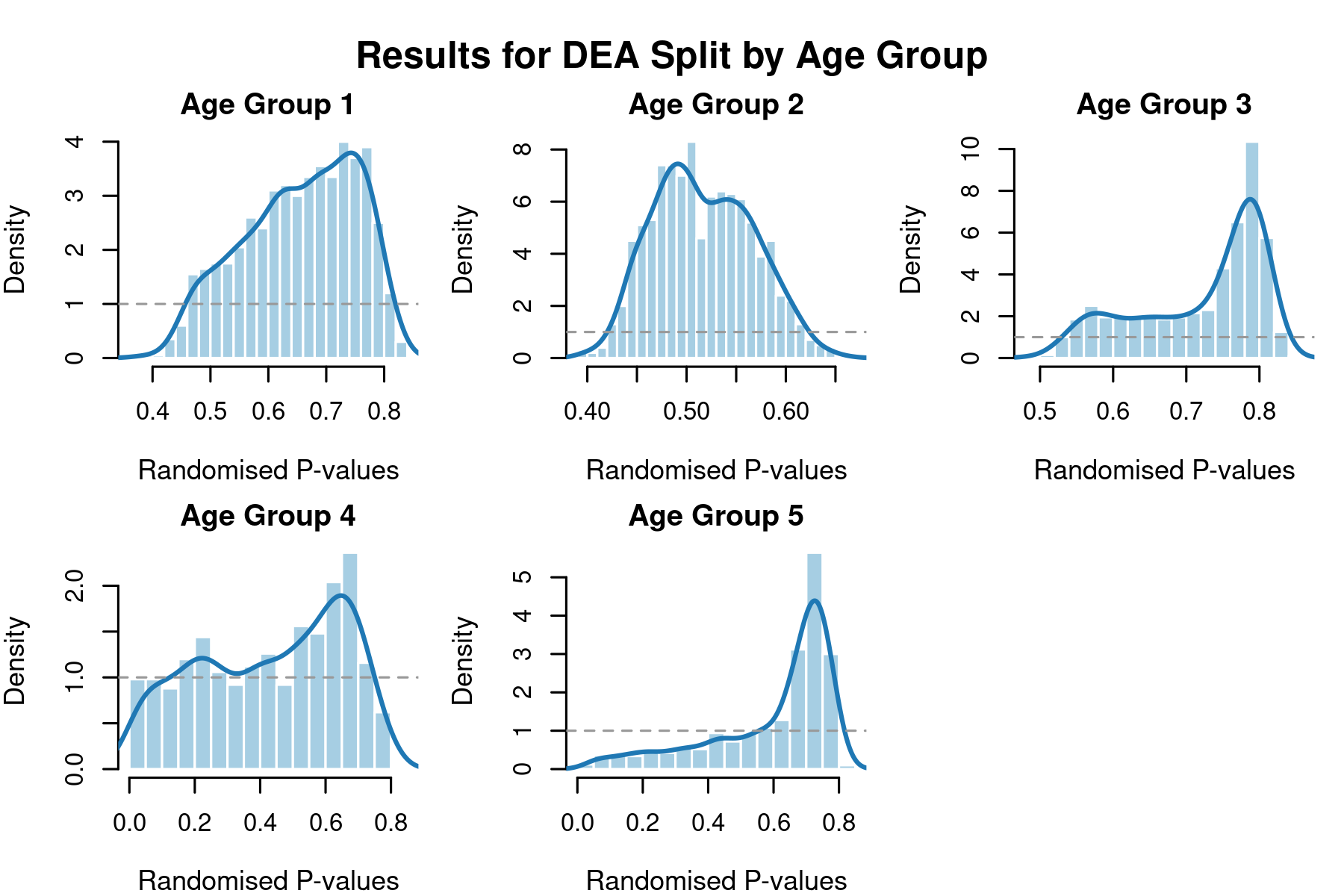}
}

\vspace{0.3cm} 

\subfloat[]{\label{fig:obs_HOS}
\includegraphics[width=0.47\linewidth]{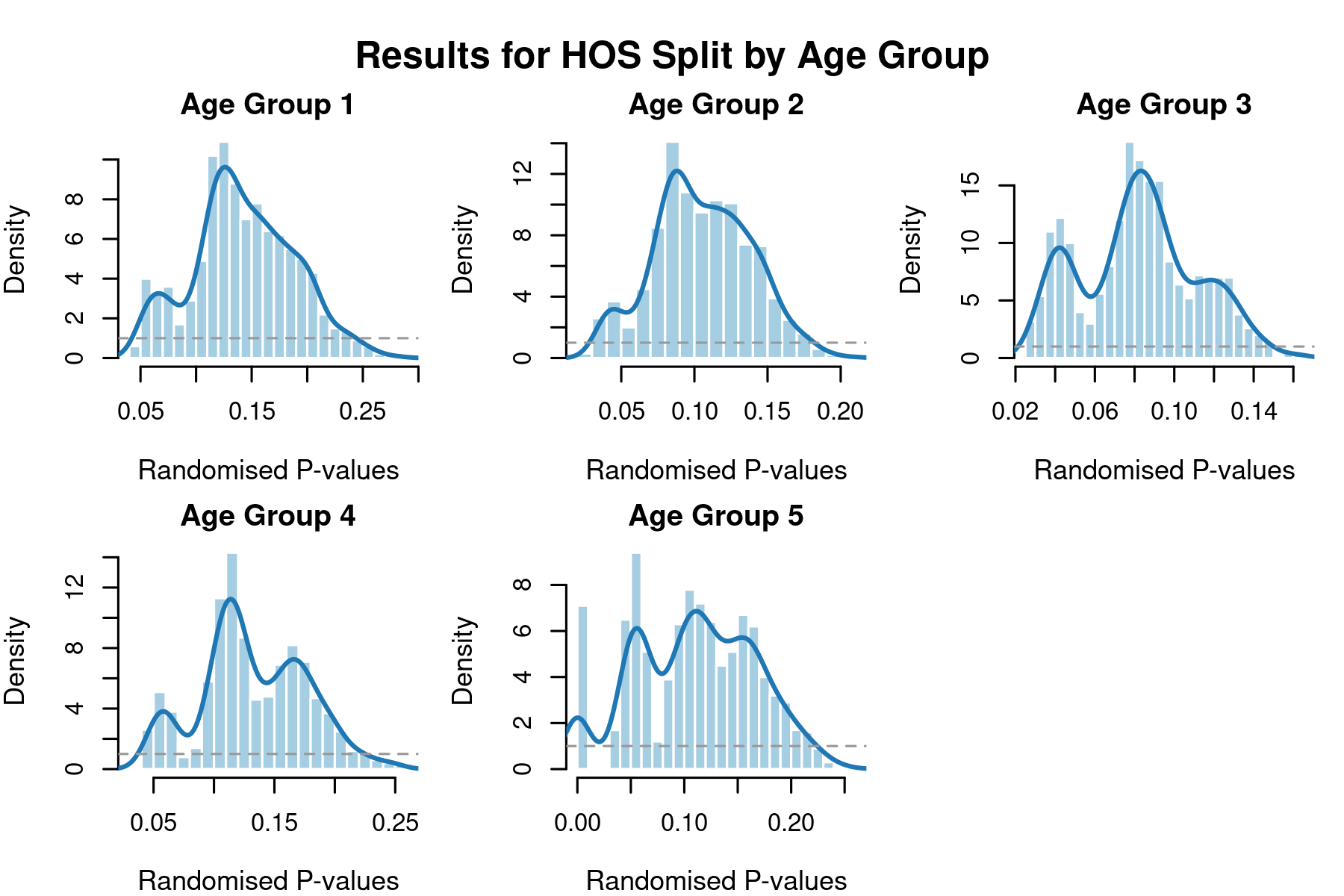}
}
\hfill
\subfloat[]{\label{fig:obs_GP}
\includegraphics[width=0.47\linewidth]{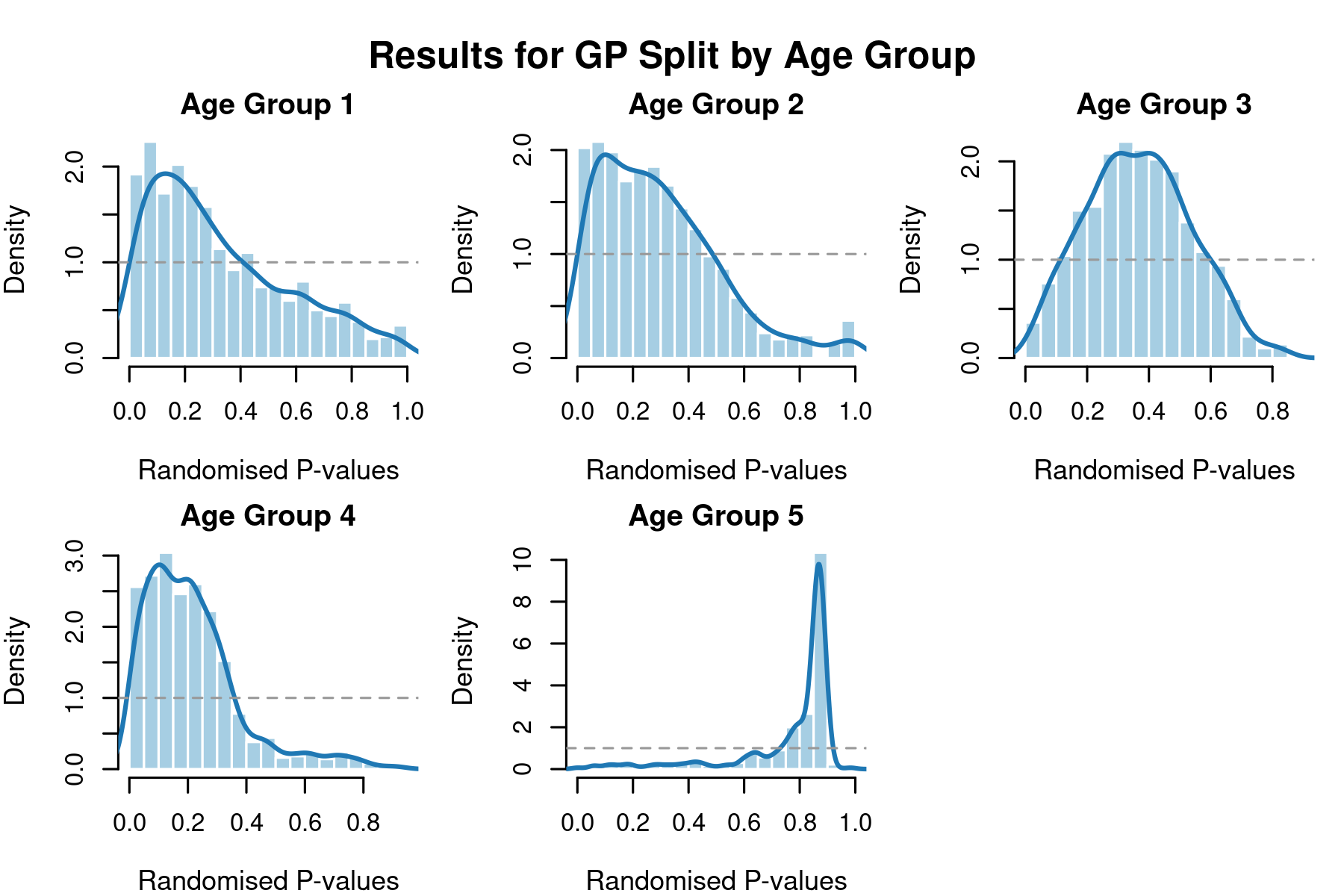}
}

\caption{The density of dependent randomised score $p$-values for each data source from the flu example. Age groups $1-5$ corresponds to $0-4$, $5-14$, $15-44$, $45-64$ and $65+$ respectively.}
\label{fig:real_distribution}
\end{figure*}

Table~\ref{tab:obs_results} presents the {combined conflict probabilities} for each data source and age group. The histograms of dependent randomised $p$-values are attached in Figure~\ref{fig:real_distribution}. From the table, we observe potential conflicts in $15+$ age groups  with the ICU data, and in the age group $15$-$44$ with the GP data. These discrepancies may indicate conflicts between the data informing the ICU and/or hospitalisation levels and the other data, or between the data informing the GP level and the remainder of the model. Alternatively, these conflicts could arise from specific modelling assumptions being inconsistent with the data, e.g. some informative priors or the aggregation of ICU data into broader age groups.

The score-based checks in the hierarchical model provide information that aligns with, and complements, the findings from the deviance-based conflict assessments (Section~\ref{criticism}). This reassures us that our approach remains effective in complex models. Moreover, it offers greater flexibility by highlighting—through the latent parameters or by selecting discrepancy measures beyond deviance—which specific components of the model may be misspecified. For instance, rather than attributing discrepancies solely to the dispersion layer or detection probabilities, the method may instead point to issues in the informative priors or the structure of the severity process as potential sources of conflict.

{
Once a discrepancy is detected, further investigation of the relevant data sources or model components may help to localise the source of conflict. The expansion used to construct the score discrepancy can then provide a natural direction for investigating its possible cause. In the present analysis, for example, the score discrepancy is constructed by expanding the binomial relationship between adjacent severity levels in the dispersion direction. For age groups showing evidence of conflict, one could therefore fit the corresponding double-binomial model with the dispersion parameter estimated rather than fixed at its binomial value. Evidence for additional dispersion would suggest that unexplained variability in the corresponding severity transition may contribute to the detected conflict. Other potential causes can be investigated analogously by introducing expansions that relax the corresponding modelling assumptions.
}

\begin{table}[htbp]
    \centering
    \caption{{Combined conflict probabilities} for the influenza example by data source and age group. Values below $0.15$ are marked blue.}
    \begin{tabular}{llccccc}
        \hline
         Sev./Age&  &$0-4$&$5-14$&$15-44$&$45-64$&$65+$ \\
        \hline
        ICU & $p_{\min}$  & {0.190} & {0.171} & \textcolor{blue}{0.112} & \textcolor{blue}{0.096} & \textcolor{blue}{0.034} \\
            & $p_\text{HCCT}$ & {0.298} & {0.240} & \textcolor{blue}{0.146} & \textcolor{blue}{0.094} & \textcolor{blue}{0.058} \\
        \hline
        DEA & $p_{\min}$  & {0.819} & {0.625} & {0.828} & {0.763} & {0.794} \\
            & $p_\text{HCCT}$ & 1.000 & 1.000 & 1.000 & {1.000} & 1.000 \\
        \hline
        HOS & $p_{\min}$  & {0.223} & {0.164} & \textcolor{blue}{0.133} & {0.209} & {0.198} \\
            & $p_\text{HCCT}$ & {0.553} & {0.283} & {0.160} & {0.457} & {0.296} \\
        \hline
        GP  & $p_{\min}$  & {0.460} & {0.468} & {0.637} & {0.332} & {0.908} \\
            & $p_\text{HCCT}$ & {0.785} & {0.771} & {1.000} & {0.364} & 1.000 \\
        \hline
    \end{tabular}
    \label{tab:obs_results}
\end{table}

\subsection{Simulation Assuming No Conflict}

We reduce the population size by a factor of $10$ (to ensure MCMC convergence at fitting stages) and simulate data from the assumed severity model, aiming to simulate data without conflict. Given that conflicts were previously detected in the elder age groups for the ICU data and in certain age groups for the HOS data, we explore the use of less informative priors for the HOS severity level. Additionally, we simulate ICU cases using a Poisson distribution as a special case of the over-dispersed model. We repeat the conflict detection procedure for the simulated dataset, and the results are summarised in Table~\ref{tab:sim_seed42} and Table \ref{tab:sim_seed123}.

\begin{table}[htbp]
    \centering
    \caption{{Combined conflict probabilities for the first simulated influenza dataset by data source and age group. Values below $0.15$ are marked in blue.}}
    \begin{tabular}{llccccc}
        \hline
        Sev./Age &  & $0-4$ & $5-14$ & $15-44$ & $45-64$ & $65+$ \\
        \hline
        ICU & $p_{\min}$  & {0.367} & {0.676} & {0.191} & {0.410} & {0.429} \\
            & $p_\text{HCCT}$ & {0.987} & 1.000 & {0.801} & {1.000} & {1.000} \\
        \hline
        DEA & $p_{\min}$  & {0.456} & {0.807} & {0.803} & {0.609} & {0.328} \\
            & $p_\text{HCCT}$ & 1.000 & 1.000 & 1.000 & 1.000 & 0.360 \\
        \hline
        HOS & $p_{\min}$  & {0.288} & {0.319} & {0.209} & {0.220} & {0.253} \\
            & $p_\text{HCCT}$ & {0.864} & {0.947} & {0.406} & {0.572} & 0.592 \\
        \hline
        GP  & $p_{\min}$  & {0.541} & {0.262} & {0.421} & {0.590} & {0.259} \\
            & $p_\text{HCCT}$ & 1.000 & {0.738} & {0.997} & {1.000} & 0.360 \\
        \hline
    \end{tabular}
    \label{tab:sim_seed42}
\end{table}

\begin{table}[htbp]
    \centering
    \caption{{Combined conflict probabilities for the second simulated influenza dataset by data source and age group. Values below $0.15$ are marked in blue.}}
    \begin{tabular}{llccccc}
        \hline
        Sev./Age & & $0-4$ & $5-14$ & $15-44$ & $45-64$ & $65+$ \\
        \hline
        ICU & $p_{\min}$  & {0.253} & {0.242} & {0.482} & {0.218} & {0.496} \\
            & $p_\text{HCCT}$ & {0.923} & 0.825 & {1.000} & {0.892} & {1.000} \\
        \hline
        GP & $p_{\min}$  & {0.670} & {0.299} & {0.515} & {0.981} & {0.746} \\
            & $p_\text{HCCT}$ & 1.000 & {0.757} & 1.000 & 1.000 & 1.000 \\
        \hline
        DEA & $p_{\min}$  & {0.720} & {0.297} & {0.978} & {0.824} & {0.643} \\
            & $p_\text{HCCT}$ & {1.000} & {0.877} & {1.000} & {1.000} & 1.000 \\
        \hline
        HOS  & $p_{\min}$  & {0.368} & {0.345} & {0.216} & {0.276} & {0.172} \\
            & $p_\text{HCCT}$ & 0.998 & {0.987} & {0.449} & {0.836} & 0.222 \\
        \hline
    \end{tabular}
    \label{tab:sim_seed123}
\end{table}

\subsection{Summary}

As our numerical experiments demonstrate, when substantial conflict is present, {combined conflict probabilities} are small. Both the minimum upper bound approach and the HCCT approach may appear somewhat conservative, as noted by the authors of these methods \citep{Yuan2012, liu2025dependent}. In practice, we recommend interpreting the {reported combined conflict probabilities} in a relative sense: a smaller value indicates weaker evidence for the absence of conflict between the given component of the model and the remaining parts. Identifying the exact components or data sources responsible for the conflict and understanding the underlying reasons often requires further investigation. {Once a potentially conflicting component has been identified, one can further introduce expansions that relax the corresponding modelling assumptions and examine whether the additional flexibility accommodates the detected conflict.}

%% file: Section6_Further.tex
\section{Discussion}
\label{sec:Discussions and Further Research Directions}



We have proposed a flexible method with a general sequential updating workflow, motivated by score discrepancy approaches for prior–data conflict detection, that enables the assessment of inconsistency between multiple data sources in evidence synthesis models.



This procedure enables conflict detection via parameter expansion at any latent node in the DAG or for any latent parameter in complex Bayesian hierarchical models. Unlike general model criticism techniques, our approach allows targeted detection of specific conflicts through freely chosen components for expansion. Our method is not strongly dependent on the overall model structure, avoiding the need to construct pivotal quantities or specify vague hyper-priors. Instead, we directly expand the prior component of interest—typically the very component we expect to absorb conflict if expanded appropriately. Our method does not rely on restrictive assumptions (e.g., normality of the posterior), nor does it require prior-predictive simulations at the data level to calibrate \( p \)-values. We have demonstrated the method's effectiveness in detecting heterogeneity across random effect groups and applied it to a more complex real-world application: a severity model for influenza infections \citep{anne2009, Presanis2013, anne2014}.

Our method can lead to conclusions that differ substantially from those obtained by node-splitting (Appendix \ref{appendix: comparison}). The latter summarises the contrasts between information from different model components, whereas our method focuses on whether the information carried by the child model induces excessive changes in the parent posterior, which would indicate conflict. In more complex examples, such as the influenza example, node-splitting may suffer from identifiability problems, whereas the score-based approach remains more robust.



Nonetheless, some limitations remain. First, the randomisation step is computationally intensive. Like cross-validation, sequential inference on sub-models using posterior draws from the parent model incurs substantial computational cost, as it requires repeated sub-model fitting. While this repetition helps preserve latent parameter uncertainty and reduces bias, it is more costly than alternatives like node-splitting, which rely on comparing just two or a few parameter estimates from partitioned models. To improve scalability, one could instead adopt efficient approximations, such as surrogate modelling and nonparametric regression (e.g. as used in Value of Information analysis \citep{Jackson2019} to avoid nested Monte Carlo simulation loops) to approximate the score discrepancy function or posterior approximations that allow the child model to be fit only once.

Second, although we have proposed some natural choices for expansion, another difficulty of using a score discrepancy is that one still needs to design test quantities capable of detecting the specific types of model misspecification of concern. Apart from providing an intuitive explanation, the analytical interpretation of the expansion parameter remains unclear beyond case-specific derivations. Further exploration of alternative discrepancy measures is therefore warranted: for example, reparametrisation through uniform latent variables \citep{Lau2014, covington2025} may give a broader view of conflicts, and links to sensitivity analysis \citep{saltelli2004sensitivity, Roos2015, Held2022, Kallioinen2024} may offer a way to explore the degree of conflict by examining the influence of priors or model components on the posterior.

Third, interpreting dependent studies remains challenging, especially in determining an appropriate way to combine quantities with a known reference distribution. The {combined conflict probabilities} can be viewed as an indication of conflict, but they may deviate substantially from the behavior of individual $p$-values obtained from predictive diagnostics. Our simulations and case studies suggest that conventional thresholds (e.g., $0.05$ or the $0.25$ suggested by \citep{Yuan2012}) are not very informative: strong conflicts tend to yield very small {combined conflict probabilities}, but even {combined conflict probabilities} for non-conflicting groups may fall below $0.25$. With larger sample sizes the power to detect smaller conflicts increases, so hypotheses are more likely to be rejected even when only minor inconsistencies are present. The Half-Cauchy combination method controls false positives at about $0.05$ but may still have limited power in some cases \citep{liu2025dependent}, a conservative trade-off between avoiding false positives and detecting true conflicts. In practice, we recommend complementing {combined conflict probabilities} with histograms of randomised $p$-values and, when signs of inconsistency emerge, examining individual model components more closely.

Finally, our method can be integrated into a general Bayesian workflow \citep{gelman2020} for modular inference and model assembly, where detecting and excluding conflicting links in frameworks such as Markov melding \citep{Goudie2019} helps prevent inconsistent combinations.

%% file: Appendix.tex
\section*{Appendix}
\label{appendix}

\setcounter{section}{0}
\setcounter{equation}{0}
\setcounter{figure}{0}
\renewcommand{\thefigure}{A\arabic{figure}}
\setcounter{table}{0}

\section{Theorems on Pivotal Quantities}
\label{Pivotal_theorems}

\subsection*{Theorem from \citep{Johnson2007}}

\textbf{Lemma.} Let $S(Y, \theta)$ denote a pivotal quantity, and suppose that $\theta_0$ is a random vector drawn from density $\pi$. Given $\theta_0$, let $Y$ denote a random vector sampled from density $f(y \mid \theta_0)$, and let $\theta_Y$ denote a parameter vector drawn from the posterior distribution on $\theta$ given $Y$. Then $S(Y, \theta_Y)$ and $S(Y, \theta_0)$ are identically distributed.

\subsection*{Theorem from \citep{Yuan2012}}

\textbf{Lemma 1.} Suppose that $d(y, \theta_0)$ is a pivotal discrepancy measure distributed according to $F$. If $\tilde{\theta}$ is drawn from the posterior distribution on $\theta$ given $y$, then $d(y, \tilde{\theta})$ is also distributed according to $F$.

In particular, this lemma applies even if $d(y, \theta_0) \equiv d(\theta_0)$: the PDM is a function of parameters only.

\section{Interpretation of Expansion in the Double Exponential Family}
\label{appendix: double}

Suppose we have a one-parameter exponential family of the format:
$$
g_{\mu,n}(y) = \exp\left[n\left\{ \eta y - \psi(\mu) \right\} \right] dG_n(y) $$ with the interpretation $y = \frac{1}{n} \sum_{i=1}^n z_i$ when $z_i \overset{\text{ind}}{\sim} g_{\mu,1}$. Here $\mu$ is the expectation parameter, $y$ is the natural statistic, $\eta$ is the natural or canonical parameter, a monotone function of $\mu$, $\psi(\mu)$ is a normalizing function, $G_n(y)$ is the carrier measure and $n$ is the sample size.

Efron \citep{Efron1986double} introduced the double exponential family:
\begin{align*}
\bar{f}_{\mu,\phi,n}(y) &= c(\mu, \phi, n) \, \phi^{1/2} \left[ g_{\mu,n}(y) \right]^{\phi} \left[ g_{y,n}(y) \right]^{1 - \phi} dG_n(y) \\
&= c(\mu, \phi, n) f_{\mu,\phi,n}(y),    
\end{align*}
allowing us to add a dispersion parameter $\phi$ to some exponential families. $c(\mu, \phi, n) \approx 1$ for the exponential family under certain conditions. $\bar{f}_{\mu,\phi,n}(y)$ itself is a two-parameter exponential family with natural parameters $(\phi\eta, \phi)$, and it approximates $g_{\mu, n\phi}(y)$. The mean of $\bar{f}_{\mu,\phi,n}(y)$ $\approx \mu$, and the variance $\approx V(\mu)/(n\phi)$. As an example, for the normal family where $y \sim \mathcal{N}(\mu, \sigma^2/n)$ with $\mu$ unknown but $\sigma^2$ a fixed and known constant, then $c(\mu, \phi, n) = 1$ and the corresponding double exponential family is $y \sim \mathcal{N}(\mu, \sigma^2/n\phi)$.

Assume $g_{y,n}(y) = \text{max}_{\mu}g_{\mu, n}(y)$. Let $V(\mu) =\mathrm{Var}_{\mu,1}\{z\} = n \,\mathrm{Var}_{\mu,n}\{y\}$. We have the KL divergence:
\begin{align*}
    &I(\mu_1, \mu_2) := \mathbb{E}_{\mu_1,1}  \log \left[ g_{\mu_1,1}(z)/{g_{\mu_2,1}(z)} \right] , \\
    &I_n(y, \mu) = nI(y, \mu) = D(y, \mu)/2.
\end{align*}

From \citep{Efron1986double}, we have score functions based on the approximate likelihood $f_{\mu,\phi,n}(y)$:
$$\frac{\partial \ell}{\partial \mu} = \frac{n \phi}{V(\mu)} (y - \mu), \quad 
\frac{\partial \ell}{\partial \phi} = \frac{1}{2\phi} -n I(y, \mu)
$$

Taking the derivative at $\phi = 1$, we have the score discrepancy
\begin{align*}
    S(\mu,\phi,n) &= \left. \frac{\partial \log \bar{f}_{\mu,\phi,n}(y)}{\partial \phi} \right|_{\phi = 1} \\
    &= \frac{1}{2} - \frac{D(y, \mu)}{2} +  \left. \frac{\partial \log c({\mu,\phi,n})}{\partial \phi} \right|_{\phi = 1}
\end{align*}
where the last term could be exactly $0$ or very small when $n$ or $\mu$ gets large, as is the case for the Binomial and Poisson distributions \citep{Efron1986double}.

With the double exponential family, expanding the dispersion parameter corresponds to calculating the deviance, which provides a natural interpretation of the expansion. In the latent space, this expansion effectively measures how extreme the deviance is between the local prior and the local likelihood, offering an alternative view of the score discrepancy as a measure of tension between prior information and observed data. This setup is a special case of the double exponential family with the dispersion parameter fixed at 1, but more generally, score discrepancy based on an expansion parameter can be used to assess conflict or lack of fit between model components.

These analytical results above also highlight a key advantage of the double exponential family: calculating score discrepancies by numerical differentiation is no longer necessary, as an explicit formula is available. This can be particularly beneficial in scenarios where computing the normalising constant is computationally expensive—such as when the denominator of the double exponential distribution becomes very large.

\section{Other Results of the Simulation Example}
\label{appendix: splitting_simulated}

\begin{figure*}[htbp]
\centering
\subfloat[No conflict between groups]{\label{fig:node_simulated_noconflict}
\centering
\includegraphics[width=0.8\linewidth]{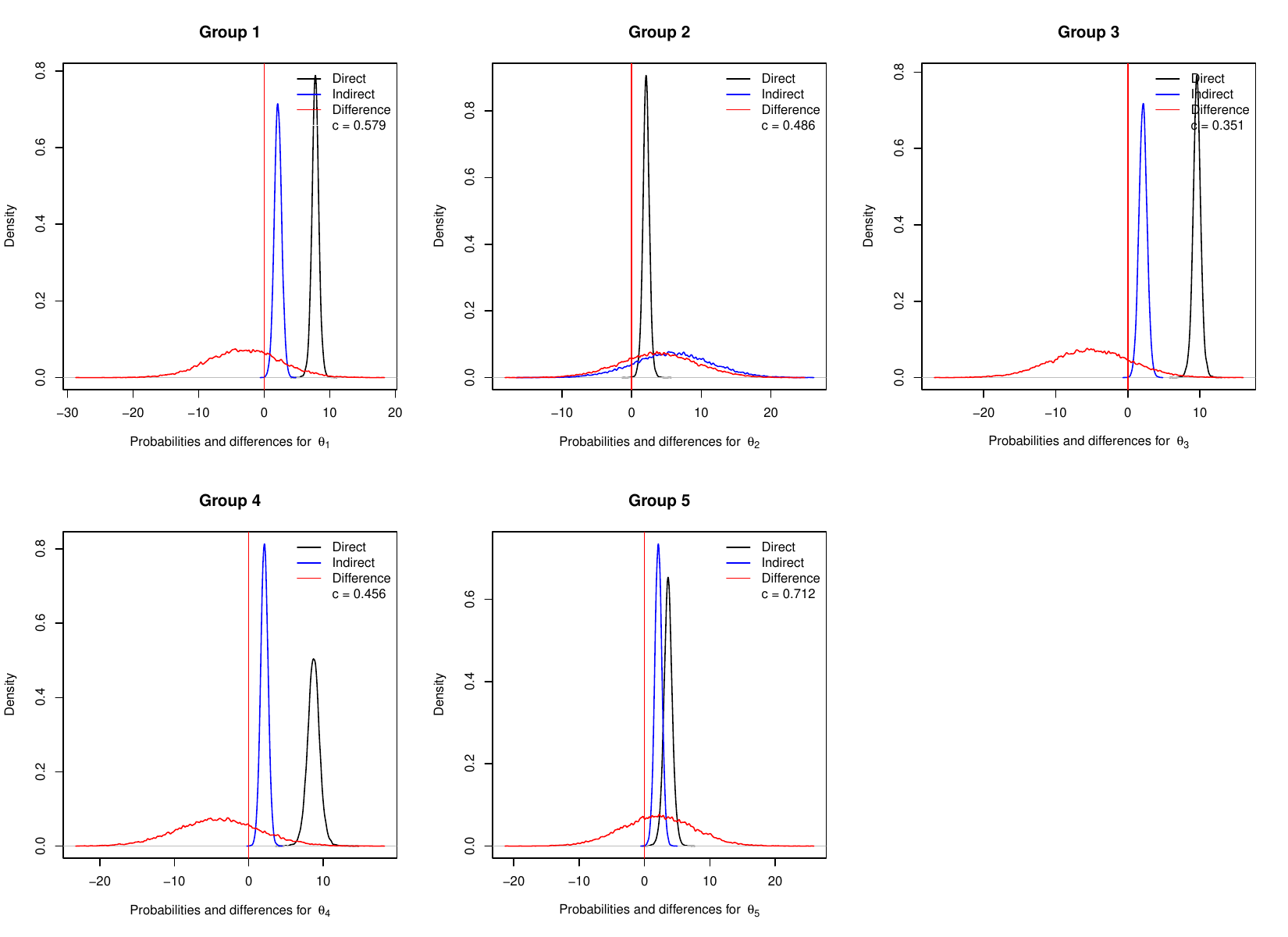}
}
\hfill
\subfloat[Conflict in Group $3$: $\theta_3 = 20$]{\label{fig:node_simulated_conflict}
\centering

\includegraphics[width=0.8\linewidth]{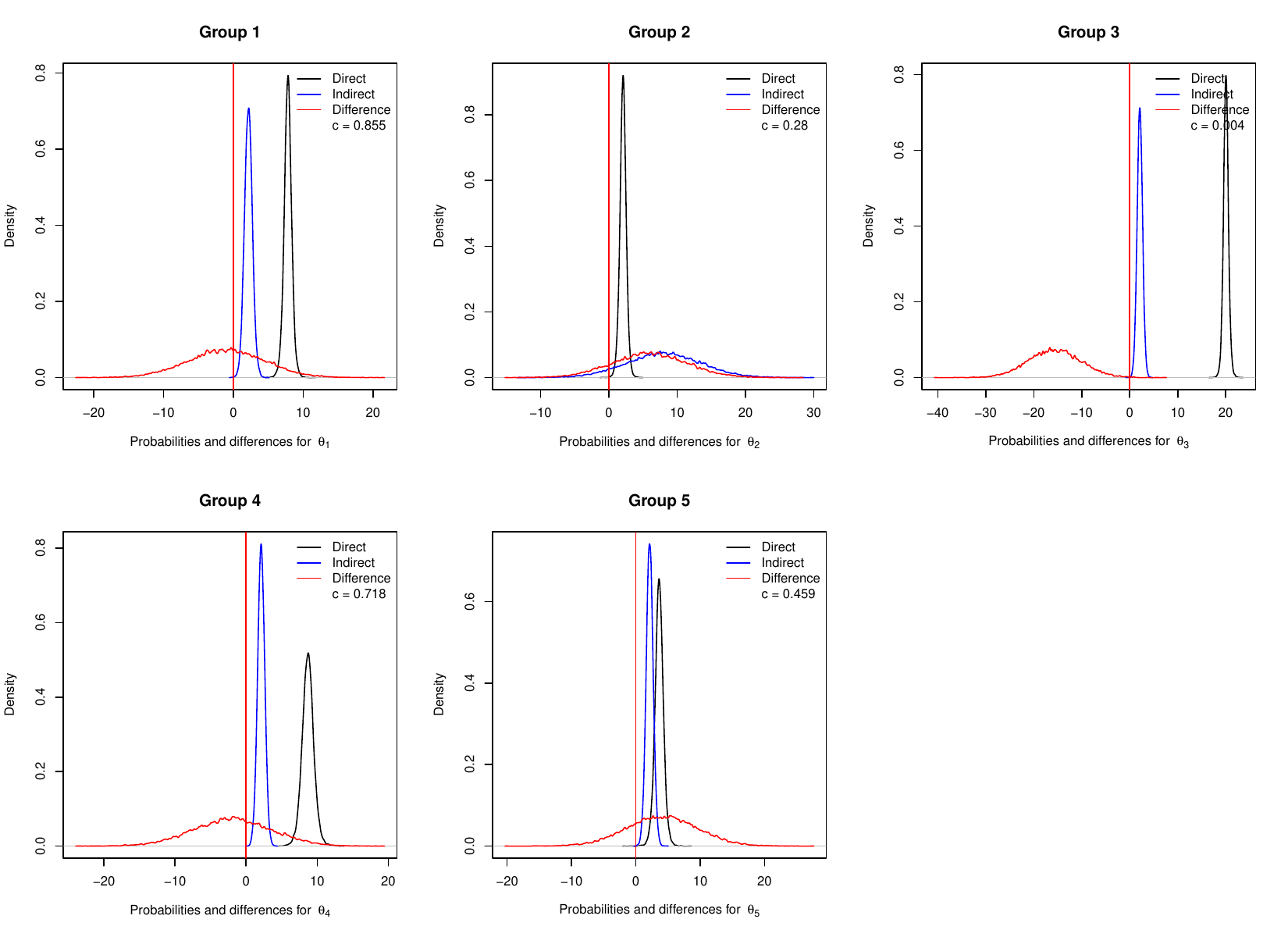}
}
\caption{Results of node-splitting in the simulation example. Posterior distributions reflecting direct (black lines) vs indirect (blue lines) evidence at $\theta_k$, the mean of each group. The conflict p-value (c), calculated as twice the proportion of MCMC samples where the difference $\theta^{\text{diff}}_k$ (red lines) greater or smaller than $0$, whichever is smaller, is given in each plot.}
\label{fig:node_simulated}
\end{figure*}
\begin{figure*}[htbp]
\centering
\subfloat[Posterior distributions and their difference for each group.]{\label{fig:node_simulated_conflict_15}
\centering
\includegraphics[width=0.8\linewidth]{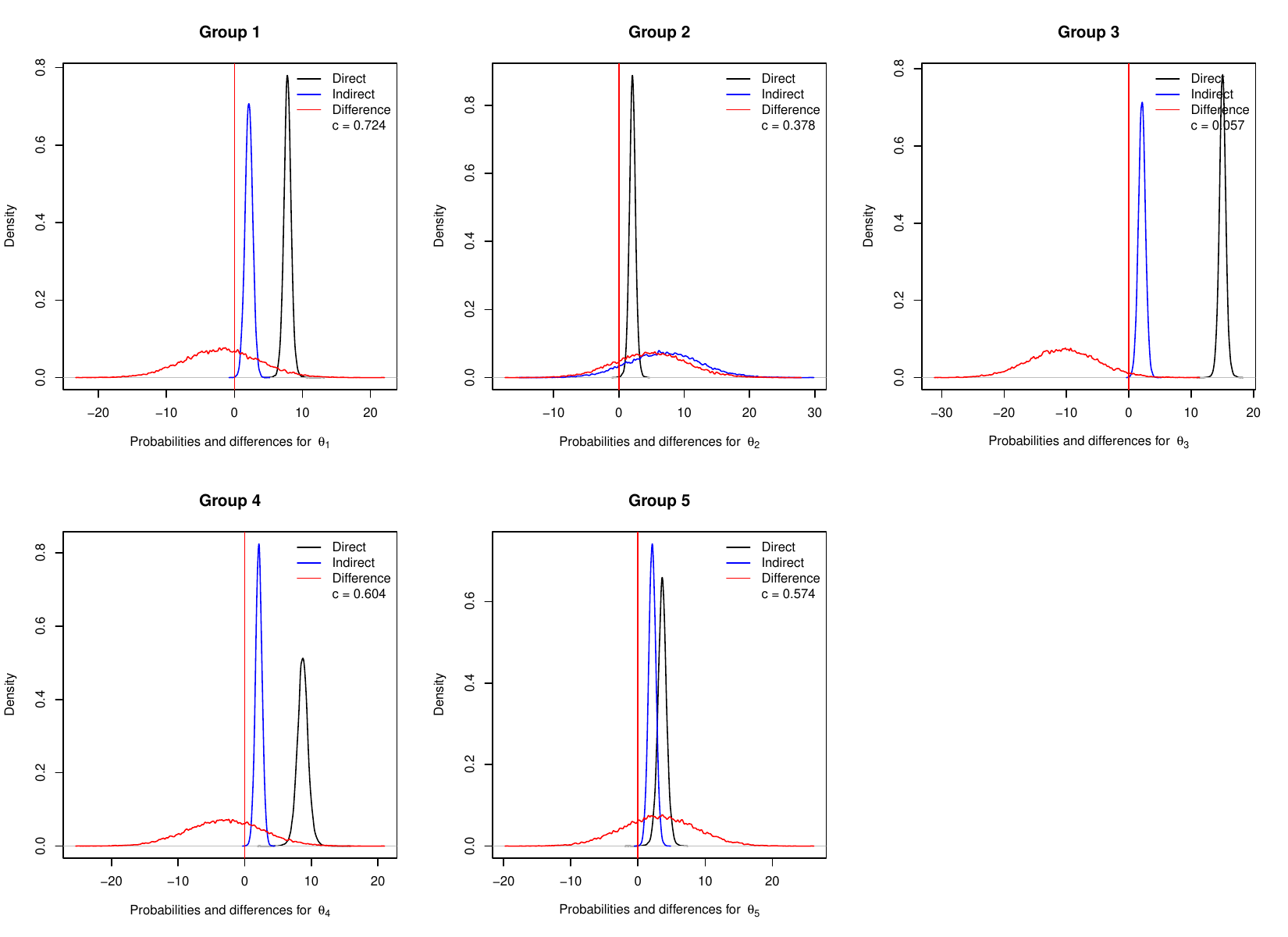}
}
\hfill
\subfloat[The distributions of randomised score $p$-value for each group.]{\label{fig:score_simulated_conflict_15}
\centering

\includegraphics[width=0.8\linewidth]{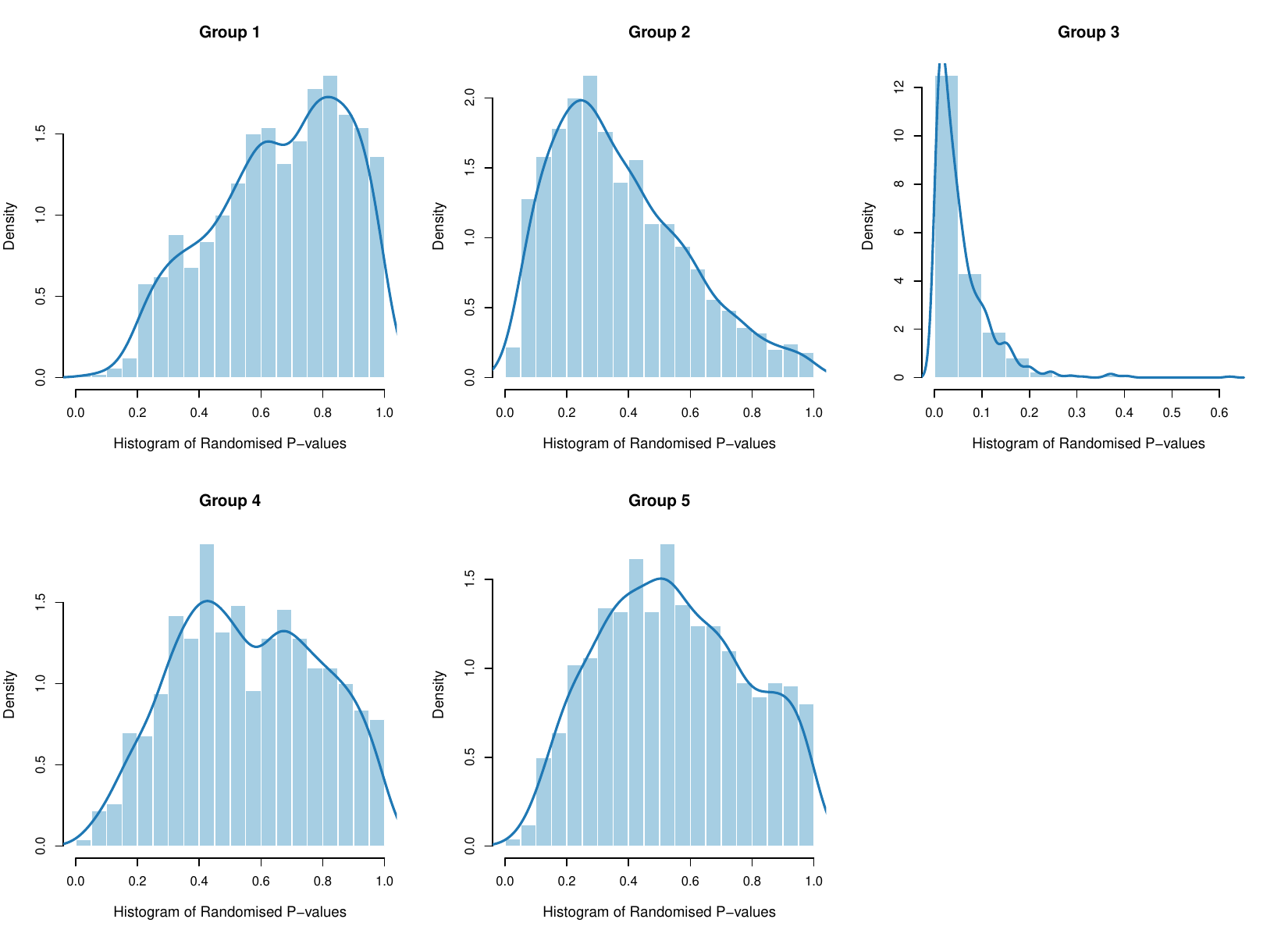}
}
\caption{Results from both methods for the simulation example when conflict is introduced by setting $\theta_3 = 15$.}
\label{fig:conflict_15}
\end{figure*}

Figure~\ref{fig:node_simulated} presents the node-splitting results for the simulation example. Figure~\ref{fig:conflict_15} displays the comparison between the node-splitting and score-discrepancy methods when a conflict is introduced by setting $\theta_3 = 15$ in the simulation example.

\section{Comparison with Conflict Measure or Node-Splitting}
\label{appendix: comparison}

In this section, we illustrate that the node-splitting approach and the score-based conflict checks quantify model conflict in different ways. Using a simple normal random effects model as a case study, we provide analytical derivations to highlight these differences. We begin by presenting the framework and analytical results from \citep{scheel2024}, followed by the derivation of corresponding results under our proposed score-based approach. Finally, we compare the two methods using distance-based measures to assess the discrepancies.

\subsection{Analytical Results of Conflict Measure (Node-Splitting) by \citep{scheel2024}}

A general exchangeable model class is given by:
\begin{align*}
Y_{i,j} &\sim p_Y(Y_{i,j} \mid \lambda_i, \boldsymbol{\phi}), \\
\lambda_i &\sim p_\lambda(\lambda_i \mid \boldsymbol{\psi}), \quad
j = 1, \dots, n_i, \quad i = 1, \dots, m
\end{align*}
with a prior distribution for $(\phi, \psi)$. We have already presented the DAG of this model in Figure $1$ of the main text.

A special but common case is the normal random effects model:
\begin{align*}
Y_{i,j} &\sim \mathcal{N}(\lambda_i, \sigma^2), \\
\lambda_i &\sim \mathcal{N}(\mu, \tau^2), \quad
j = 1, \dots, n, \quad i = 1, \dots, m.
\end{align*}
with a prior on $(\sigma^2, \mu, \tau^2)$. For illustration, we consider a further simplified case assuming known variances and a flat prior for $\mu$: $ \sigma^2 = \sigma_0^2$, $\tau^2 = \tau_0^2$, $\pi(\mu) = 1$.

\subsubsection{Information Contributions}
A large class of Bayesian hierarchical models can be represented using DAGs. By the factorisation property, the joint probability distribution for all variables is given by:
$$
    p(\mathbf{Y},\boldsymbol{\theta}) = \prod_{Y \in \mathbf{Y}} p(Y | \text{Pa}(y)) \prod_{\theta \in \boldsymbol{\theta}} p(\theta | \text{Pa}(\theta))
$$
where $\text{Pa}()$ are the parent nodes of a node, and $\text{Ch}()$ denotes child nodes correspondingly.

For each $\lambda_i$ in the exchangeable model, the conditional probability is:
$$
    p(\lambda_i | (y,\theta)_{-\lambda_i}) \propto p(\lambda_i | \text{Pa}(\lambda_i)) \prod_{\gamma \in \text{Ch}(\lambda_i)} p(\gamma | \text{Pa}(\gamma)).
$$
Intuitively, each node in a DAG receives information from its parent and child nodes. Hence, the first term can be thought of as the local prior information contribution and the second term involving child nodes of $\lambda_i$ is the local likelihood information source \citep{scheel2011graphical}. Formally, we define the prior and likelihood information contributions \citep{dahl2007robust} respectively for $\lambda_i$ (essentially reorganisation of the above conditional probability):
$$
    f_p(\lambda_i; \beta_p) = p(\lambda_i | \beta_p), \quad f_c(\lambda_i; \mathbf{\beta}_c) \propto \prod_{\gamma \in \text{Ch}(\lambda_i)} p(\gamma | \text{Pa}(\gamma))
$$ where $\beta_p = \text{Pa}(\lambda_i)$ and $\beta_c = \bigcup_{\gamma \in \text{Ch}(\lambda_i)} (\{\gamma\} \cup \text{Pa}(\gamma)) - \{\lambda_i\}$ (note $\beta_c$ may contain data nodes). We assume $f_c$ is a density.

Integrating out latent parameters, the integrated information contributions (iic) defined in \citep{gasemyr2016uniformity} are given by:
\begin{align*}
g_p(\lambda_i) 
&= \int f_p(\lambda_i; \beta_p)\, \pi(\beta_p \mid \mathbf{y}_p)\, d\beta_p, \\
g_c(\lambda_i) 
&= \int f_c(\lambda_i; \beta_c)\, \pi(\beta_c \mid \mathbf{y}_c)\, d\beta_c.
\end{align*}
depending on suitable data-splitting: $\mathbf{y}_c = \mathbf{y} \cap \text{Desc}(\lambda_i)$ and $ \mathbf{y}_p = \mathbf{y} - \mathbf{y}_c.$ By \citep{gasemyr2016uniformity}, the densities $g_p$ and $g_c$ can in general be expressed as posterior densities given data $\mathbf{y}_p$ and $\mathbf{y}_c$ respectively, the latter using the improper or vague prior density $\pi(\lambda_i) = 1$, independently of the co-parents. This means that the empirical CDF of $G_p$ and $G_c$ can be obtained directly as posterior samples e.g.,~from MCMC outputs, which aligns with the node-splitting framework suggested by \citep{Presanis2013}.

For the simple normal example, the explicit forms of iic are:
\begin{align*}
    f_c(\lambda_i) &= g_c(\lambda_i) = \mathcal{N}(\lambda_i; \bar{Y}_i, \sigma_0^2/n), \\
    f_p(\lambda_i) &= \mathcal{N}(\lambda_i; \mu, \tau_0^2), \\
    g_p(\lambda_i) &= \mathcal{N}(\lambda_i; \bar{Y}_{-i}, (m/(m-1)) \tau_0^2 + \sigma_0^2/(n(m-1)))
\end{align*}
where $\bar{Y}_{-i} = \frac{1}{n(m-1)} \sum_{k \neq i} \sum_{j=1}^{n} Y_{k,j}$.

\subsubsection{Conflict Measure (Conflict $p$-value)}

For a given pair \( G_p, G_c \) of iic distributions, let \( \lambda^*_p \sim G_p\) and \( \lambda^*_c \sim G_c\) be independent samples. Let \( G \) be the CDF for \( \delta = \lambda^*_p - \lambda^*_c \). Define the conflict measures
\[
c^3_{\lambda_i} = 1 - 2 \min(G(0), 1-G(0)), \quad c^4_{\lambda_i} = P_G(g(\delta) > g(0)).
\]
as in \citep{gasemyr2009extensions}, which are consistent with the conflict $p$-values of \citep{Presanis2013}. The latter is based on the tail areas of $g$.

For the simple normal example, it follows from the previous results that
\begin{align*}
    g(\delta)& = \mathcal{N}\left(\delta; \bar{Y}_{-i} - \bar{Y}_i, \frac{m}{m - 1} \left(\tau_0^2 + \frac{\sigma_0^2}{n}\right) \right), \text{ and}\\
    g(0) &= \mathcal{N}\left(0; \bar{Y}_i - \bar{Y}_{-i}, \frac{m}{m - 1} \left(\tau_0^2 + \frac{\sigma_0^2}{n}\right) \right).
\end{align*}

Since the variable \( \bar{Y}_{-i} - \bar{Y}_i \) is normally distributed with expectation $0$ and variance \( \frac{m}{m - 1} (\tau_0^2 + \sigma_0^2/n) \), it follows that \( c^{3}_{\lambda_i} \) and \( c^{4}_{\lambda_i} \) are both {uniformly distributed pre-experimentally and equivalent} in this example. $c^3$ and $c^4$ are equivalent and meaningful under many cases, e.g., for symmetric uni-modal distributions, and are well-calibrated even in a variety of non-Gaussian situations \citep{gasemyr2016uniformity}. Summarising information from partitioned sub-models, cross-validatory measures also generally tend to be well-calibrated \citep{Marshall2007}.

\subsection{Analytical Results of Score-Based Conflict Checks}

In the simple normal example, for $\lambda_i$ we have the following sequential analysis: First we derive a posterior distribution for \( \mu \) given \( Y_{-i} \) with the full model and draw samples \( \tilde{\mu}  \sim  p(\mu | \mathbf{Y}_{-i}) \). For each single draw \( \tilde{\mu} \): we obtain a single posterior \( \tilde{\lambda}_i \sim p(\lambda_i | \tilde{\mu}, Y_i) \) from the submodel of group $i$: $Y_{i} \sim \mathcal{N}(Y_{i} | \lambda_i, \sigma^2_0)$, $\lambda_i \sim \mathcal{N}(\lambda_i | \tilde{\mu}, \tau^2_0)$; then carry out conflict check for \(\lambda_i\) with expansion on standard deviation $\mathcal{N}(\lambda_i | \mu, \tau^2_0\alpha) $. We compare
        $
        \left. \frac{d}{d\alpha} \log \mathcal{N}(\tilde{\lambda}_i | \tilde{\mu}, \tau^2_0\alpha) \right|_{\alpha = 1}
        $ to
        $
        \left. \frac{d}{d\alpha} \log \mathcal{N}(\lambda_i | \tilde{\mu}, \tau^2_0\alpha) \right|_{\alpha = 1}
        $
        where \( \lambda_i \sim \mathcal{N}(\lambda_i | \tilde{\mu}, \tau^2_0) \), giving randomised (one-sided) score p-values.

For \( \tilde{\mu}  \sim  p(\mu | Y_{-i}) \) and each single draw $ \tilde{\lambda}_i$, we have
\begin{align*}
    &\tilde{\mu} \sim \mathcal{N}\left(\mu; \bar{Y}_{-i}, \frac{1}{m-1}( \tau_0^2 + \sigma_0^2/n)\right), \\
    &\tilde{\lambda}_i \mid \tilde{\mu}, Y_i \sim \mathcal{N}\left(\frac{1}{M}\left({\frac{n}{\sigma^2_0}\bar{Y}_i+\frac{1}{\tau^2_0} \tilde{\mu}}\right), \frac{1}{M}\right)
\end{align*}where $M:= \frac{n}{\sigma^2_0}+\frac{1}{\tau^2_0}$.

Notice
\begin{equation*}
    \left. \frac{d}{d\alpha} \log \mathcal{N}(x | \mu, \sigma^2\alpha) \right|_{\alpha = 1} \dot=  (x-\mu)^2/ \sigma^2.
\end{equation*} Thus in the next step, we are in fact comparing the absolute value of
$X = (\tilde{\lambda}_i-\tilde{\mu})/\tau_0$ to $Z= (\lambda_i-\tilde{\mu})/\tau_0$ where \( \lambda_i \sim \mathcal{N}(\lambda_i | \tilde{\mu}, \tau^2_0) \).
\begin{equation*}
X \mid \tilde{\mu} \sim \mathcal{N}\left( \frac{1}{M\tau_0}\frac{n}{\sigma^2_0}(\bar{Y}_i- \tilde{\mu}),\frac{1}{M\tau^2_0} \right),
\end{equation*} and $Z$ has a standard normal distribution.
We write
\begin{equation*}
\begin{split}
    X \mid \tilde{\mu} = \frac{n}{\sigma^2_0}\frac{1}{M\tau_0}(\bar{Y}_i-\mu') + \frac{1}{\sqrt{M}\tau_0} \varepsilon, \quad \varepsilon \sim \mathcal{N}(0,1); \\
    \tilde{\mu} = \bar{Y}_{-i} + \frac{1}{\sqrt{m-1}}{\sqrt{\tau_0^2 + \frac{\sigma_0^2}{n}}} \eta, \quad \eta \sim \mathcal{N}(0,1)
\end{split}
\end{equation*}
Integrating out $\tilde{\mu}$, we get
\begin{align*}
X 
&\sim 
\mathcal{N}\!\left(
    \frac{\bar{Y}_i - \bar{Y}_{-i}}
         {\sigma_0^2 / (n\tau_0) + \tau_0},
    \frac{\sigma_0^2 + n\tau_0^2 / (m - 1)}
         {n\tau_0^2 + \sigma_0^2}
\right) \\
&= 
\mathcal{N}\!\left(
    \frac{\bar{Y}_i - \bar{Y}_{-i}}{K},
    \frac{\sigma_0^2 / (n\tau_0) + \tau_0 / (m - 1)}{K}
\right).
\end{align*}
where $K:= \sigma_0^2/n\tau_0 + \tau_0$. This quantity $X$ is compared with the standard normal distribution and $p$-values are obtained from the two-sided tail areas. We call $X$ the scaled difference between groups from score-based conflict checks.

\subsection{Comparison of Two Methods}

From the distinct forms of the $p$-values above, we should indeed expect different $p$-values from the different methods, especially in complicated examples. Not to mention that our score discrepancy approach is also based on combining multiple dependent \( p \)-values. For our method, the randomisation stage in implementation offers a straightforward approach to calibrating \( p \)-values; however, it comes at the cost of increased computational burden and the need for interpreting dependent studies.

\begin{figure}[h]
    \centering
    \includegraphics[width=0.6\linewidth]{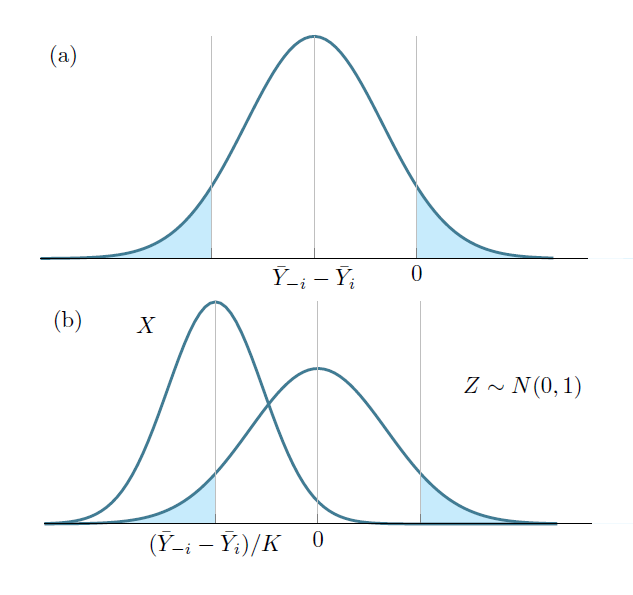}
    \caption{(a) shows how we obtain $p$-values from the node-splitting approach for the simple example, and (b) illustrates that with the score-based check.}
    \label{fig:comparison}
\end{figure}

For the simple example, the node splitting approach is to compare $0$ to the distribution $\mathcal{N}\left( \bar{Y}_{-i} - \bar{Y}_i, \frac{m}{m - 1} \left(\tau_0^2 + \frac{\sigma_0^2}{n}\right) \right)$ and calculate the two-sided tail areas. The score-based check is to compare the scaled difference $X$ to the standard normal and calculate the two-sided tail areas where $X$ itself has a distribution centering at the scaled difference $\mathcal{N}\left((\bar{Y}_i-\bar{Y}_{-i}) \frac{1}{K},\left( \frac{\sigma_0^2}{n\tau_0}+ \frac{\tau_0}{m-1} \right) \frac{1}{K}\right)$. We visiualise this difference in Figure~\ref{fig:comparison}.

\section{Influenza Model Extensions}
\label{Flu_extension}

\subsection{Introducing an extra GP consultation level}

In the third wave severity analyses of \citep{anne2014} for the UK, the number of GP consultations specific to the pandemic A/H1N1 strain was estimated using a joint regression model of GP consultations for influenza-like-illness (ILI) and sentinel testing data on proportions of nasopharyngeal swabs testing positive for A/H1N1pdm from a subset of these consultations.

This estimate was scaled up to an estimate of the number of symptomatic infections by combining it with an estimate of the probability asymptomatic infection leads to a GP consultation from the second half of the second wave, from the transmission model of \citep{Birrell2011}. This combination occurred before inclusion of the estimate in the third wave severity model, as a likelihood term.

An alternative is to incorporate the pandemic strain GP consultation rate estimate from the test postivity/GP model directly in the severity model, by introducing an extra severity level to represent GP consultation, between the symptomatic and hospitalisation levels. The posterior mean of the estimated number of pandemic strain GP consultations is incorporated as a likelihood term, in such a way to reflect the posterior uncertainty in the estimate, as detailed in Section~\ref{sec: over-dispersion}. The posterior probability of a GP consultation given symptomatic infection from \citep{Birrell2011} is also incorporated as a prior for the corresponding probability in the severity model, $p_{a,\text{GP} \mid \text{SYM}}$.

This alternative formulation allows feedback from all the other evidence (data and priors) in the severity model to the estimates of the number of GP consultations and $p_{a,\text{GP} \mid \text{SYM}}$ from the test positivity/GP and transmission models.

\subsection{Over-dispersion}
\label{sec: over-dispersion}

We also consider allowing over-dispersion in the data to resolve the poor-mixing issue when we introduce stochasticity between severity levels. 



We include over-dispersion together with dectection probabilities by considering the count observations to be realisations of the negative binomial distribution, with dispersion parameter $\psi_{a,\ell}$, for each $\ell \in \{\text{GP},\text{HOS},\text{ICU},$ $ \text{DEA}\}$:
\begin{align}
y_{a,\ell} & \sim \text{Negative-Binomial}(\psi_{a,\ell}, r_{a,\ell}) \nonumber\\
r_{a,\ell} & = N^{(u)}_{a,\ell} \psi_{a,\ell} / (1 - \psi_{a,\ell}) \nonumber\\
\psi_{a,\ell} & \sim \text{Beta}(\alpha^{(\psi)}_{a,\ell}, \beta^{(\psi)}_{a,\ell})
\label{eqn:over-dispersion}
\end{align}
Here, $r_{a,\ell}$ is an intermediate size parameter in the standard $(r, p)$ form of the negative binomial, with $p = \psi_{a,\ell}$ and the mean fixed at $N^{(u)}_{a,\ell}$. So that the observations have mean $\mathbb{E}(Y_{a,\ell}) = N^{(u)}_{a,\ell}$, variance $\text{Var}(Y_{a,\ell}) = N^{(u)}_{a,\ell} / \psi_{a,\ell}$ and $N^{(u)}_{a,\ell}$ refers to the (potentially) under-ascertained version of the number of infections at level $\ell$. $N^{(u)}_{a,\ell}$ is therefore a function of the actual number $N_{a,\ell}$ and the corresponding detection probability $d_{a,\ell}$:
\begin{align}
\log(N^{(u)}_{a,\ell}) & = \log(N_{a,\ell}) + \log(d_{a,\ell}) \nonumber\\
d_{a,\ell} & \sim \text{Beta}(\alpha^{(d)}_{\ell},\beta^{(d)}_{\ell})
\label{eqn:under-ascertainment}
\end{align} The over-dispersion with detection probabilities is also shown in Figure~\ref{fig:flu_DAG_extension}.
Note that the ICU immigration-death process submodel provides estimates only for two broad age groups: $<15, 15+$. Consequently, the quantities $N^{(u)}_{a,\ell}$ need to be aggregated to match this age stratification.

From the same source as the hospitalisation count data---the sentinel hospital dataset---we also obtain estimates of the conditional probability of ICU admission given hospitalisation ($p_{a, \text{ICU} \mid \text{HOS}}$). We incorporate this information by modelling the observed number of ICU admissions ($y_p$) as a binomial random variable with size equal to the observed number of hospitalisations ($n_p$).

We specify vaguely informative priors for the dispersion parameters $\psi_{a,\text{HOS}}$ and $\psi_{a,\text{DEA}}$ to allow small over-dispersion in the the hospitalisation and death data. In contrast, we assign informative priors to $\psi_{a,\text{GP}}$ and $\psi_{a,\text{ICU}}$, derived from the test positivity/GP model and the ICU submodel (see \citep{anne2014}), respectively. These priors are intended to reflect the posterior uncertainty captured in those source models. Therefore, in this example, we ultimately use a hierarchical beta-binomial model with over-dispersion to check for conflict for this example. We incorporate both over-dispersion and detection probabilities to account for under-ascertainement.